\UseRawInputEncoding
\pdfoutput=1
\documentclass[11pt]{article}

\usepackage{selinput}
\usepackage{amsmath}
\usepackage{xspace}
\usepackage{microtype}
\usepackage{xcolor}
\usepackage{adjustbox}
\usepackage{graphicx}
\usepackage{subfigure}
\usepackage{booktabs}
\usepackage[skip=0pt]{caption}
\usepackage[hyphens]{url}
\usepackage{hyperref}
\usepackage{array}
\usepackage{float}
\usepackage{listings}
\usepackage{titlesec} 
\usepackage{times}
\usepackage{latexsym}
\usepackage[T1]{fontenc}
\usepackage{inconsolata}
\usepackage{xcolor}
\definecolor{DarkGreen}{RGB}{0,100,0} 
\definecolor{DarkRed}{RGB}{178,34,34} 
\usepackage[final]{ACL2023}
\usepackage{comment}
\usepackage{multirow}
\usepackage[hang,flushmargin]{footmisc}
\usepackage{amssymb}
\usepackage{amsmath}

\usepackage[ruled,vlined]{algorithm2e} 
\usepackage{algorithmic}
\lstdefinelanguage{Prompt}{
    basicstyle=\ttfamily\small,
    frame=single,
    backgroundcolor=\color{gray!10}, 
    breaklines=true,                
    keywordstyle=\color{blue}\bfseries,    
    stringstyle=\color{red},             
    commentstyle=\color{green!50!black},  
    morekeywords={CREATE, RETURN, SELECT, WHERE, AS, MATCH, WHEN, THEN, ELSE, END, FROM, ORDER, BY, LIMIT},
    literate={\{}{{\textcolor{cyan}{\{}}}{1} 
             {\}}{{\textcolor{cyan}{\}}}}{1} 
             {self.table}{{\textcolor{magenta}{self.table}}}{10} 
             {self.question}{{\textcolor{magenta}{self.question}}}{13} 
             {self.description}{{\textcolor{magenta}{self.description}}}{16} 
             {self.table.columns}{{\textcolor{magenta}{self.table.columns}}}{18} 
             {df.column}{{\textcolor{magenta}{df.column}}}{9} 
             {self.plan}{{\textcolor{magenta}{self.plan}}}{9} 
             {self.plan}{{\textcolor{magenta}{self.plan}}}{9} 
             {step.prompt}{{\textcolor{magenta}{step.prompt}}}{11} 
             {self.name}{{\textcolor{magenta}{self.name}}}{9}
             {table_title}{{\textcolor{blue}{table\_title}}}{11}
             {table_description}{{\textcolor{blue}{table\_description}}}{17}
             {table_summary}{{\textcolor{blue}{table\_summary}}}{13}
             {col_desc}{{\textcolor{blue}{col\_desc}}}{8}
             {col_format}{{\textcolor{blue}{col\_format}}}{10}
             {entities}{{\textcolor{blue}{entities}}}{8}
             {actions}{{\textcolor{blue}{actions}}}{7}
             {details}{{\textcolor{blue}{details}}}{7}
             {events}{{\textcolor{blue}{events}}}{6}
             {timeline}{{\textcolor{blue}{timeline}}}{8}
             {topic}{{\textcolor{blue}{topic}}}{5}
             {source_name}{{\textcolor{red}{source\_name}}}{11}
             {table_content}{{\textcolor{red}{table\_content}}}{13}
             {doc_title}{{\textcolor{red}{doc\_title}}}{9}
             {doc_content}{{\textcolor{red}{doc\_content}}}{11}
             {Example}{{\textbf{Example}}}{7}
             {MySQL_Code_Generation:}{{\textbf{MySQL Code Generation:}}}{22} 
             {Instructions:}{{\textbf{Instructions:}}}{12} 
             {Step_1}{{\textbf{Step 1}}}{6}
             {Step_2}{{\textbf{Step 2}}}{6}
             {Step_3}{{\textbf{Step 3}}}{6}
             {Step_4}{{\textbf{Step 4}}}{6}
             {Step_5}{{\textbf{Step 5}}}{6}
             {Step_6}{{\textbf{Step 6}}}{6}
             {Question_:}{{\textbf{Question:}}}{9}
             {LLM_Step}{{\textbf{LLM Step}}}{8}                               
             {SQL_Step}{{\textbf{SQL Step}}}{8}
             {Champion_}{{\textcolor{red}{'\%Champion\%'}}}{13} 
        {Win__}{{\textcolor{red}{'Win'}}}{5}
        {Round__}{{\textcolor{red}{'\%1st Round\%'}}}{14}
        {No_Win}{{\textcolor{red}{'No Win'}}}{8}
        {1936_}{{\textcolor{red}{'1936'}}}{6}
}

\begin{document}
\newcommand{\methodname}{{\sc SAGE}\xspace}
\newcommand{\agentname}{{\sc SPARK}\xspace}
\title{\methodname: Structure Aware Graph Expansion for Retrieval of Heterogeneous Data}

\author{
Prasham Titiya$^{1*}$,
Rohit Khoja$^{1*}$,
Tomer Wolfson$^{2\dagger}$,
Vivek Gupta$^{1\dagger}$\\
\textbf{Dan Roth}$^{2}$, 
\\[4pt]
$^{1}$ Arizona State University \quad
$^{2}$ University of Pennsylvania
\\[2pt]
{\small $^{*}$Equal contribution (co-first authors)\quad
$^{\dagger}$Equal contribution (co-second authors)}
}

\maketitle
\begin{abstract}
Retrieval-augmented question answering over heterogeneous corpora requires \emph{connected} evidence across text, tables, and graph nodes. While entity-level knowledge graphs support structured access, they are costly to construct and maintain, and inefficient to traverse at query time. In contrast, standard retriever-reader pipelines use flat similarity search over independently chunked text, missing multi-hop evidence chains across modalities. We propose \methodname (\textbf{S}tructure \textbf{A}ware \textbf{G}raph \textbf{E}xpansion) framework that (i) constructs a chunk-level graph \emph{offline} using metadata-driven similarities with percentile-based pruning, and (ii) performs \emph{online} retrieval by running an initial baseline retriever to obtain $k$ seed chunks, expanding first-hop neighbors, and then filtering the neighbors using dense+sparse retrieval, selecting $k'$ additional chunks. We instantiate the initial retriever using hybrid dense+sparse retrieval for implicit cross-modal corpora and \agentname (\textbf{S}tructure Aware \textbf{P}lanning \textbf{A}gent for \textbf{R}etrieval over \textbf{K}nowledge Graphs) an agentic retriever for explicit schema graphs. On OTT-QA and STaRK, \methodname improves retrieval recall by 5.7 and 8.5 points over baselines. \noindent\footnote{ Code available at \url{https://github.com/CoRAL-ASU/sage}}
\end{abstract}

\section{Introduction}

Retrieval-augmented generation (RAG) \cite{lewis2020retrieval} grounds large language models in external knowledge and has become a standard paradigm for question answering over domain-specific corpora. Most RAG systems rely on flat retrieval, where independently indexed text chunks are retrieved using sparse, dense, or hybrid similarity search \cite{mandikal2024sparse,arabzadeh2021predicting,zhang2024efficient,chen2022salient}. While effective for simple text-centric queries, this design struggles with complex reasoning over heterogeneous data that includes documents, tables, and structured metadata.

\begin{figure}[h]
\centering
\includegraphics[width=1.0\linewidth]{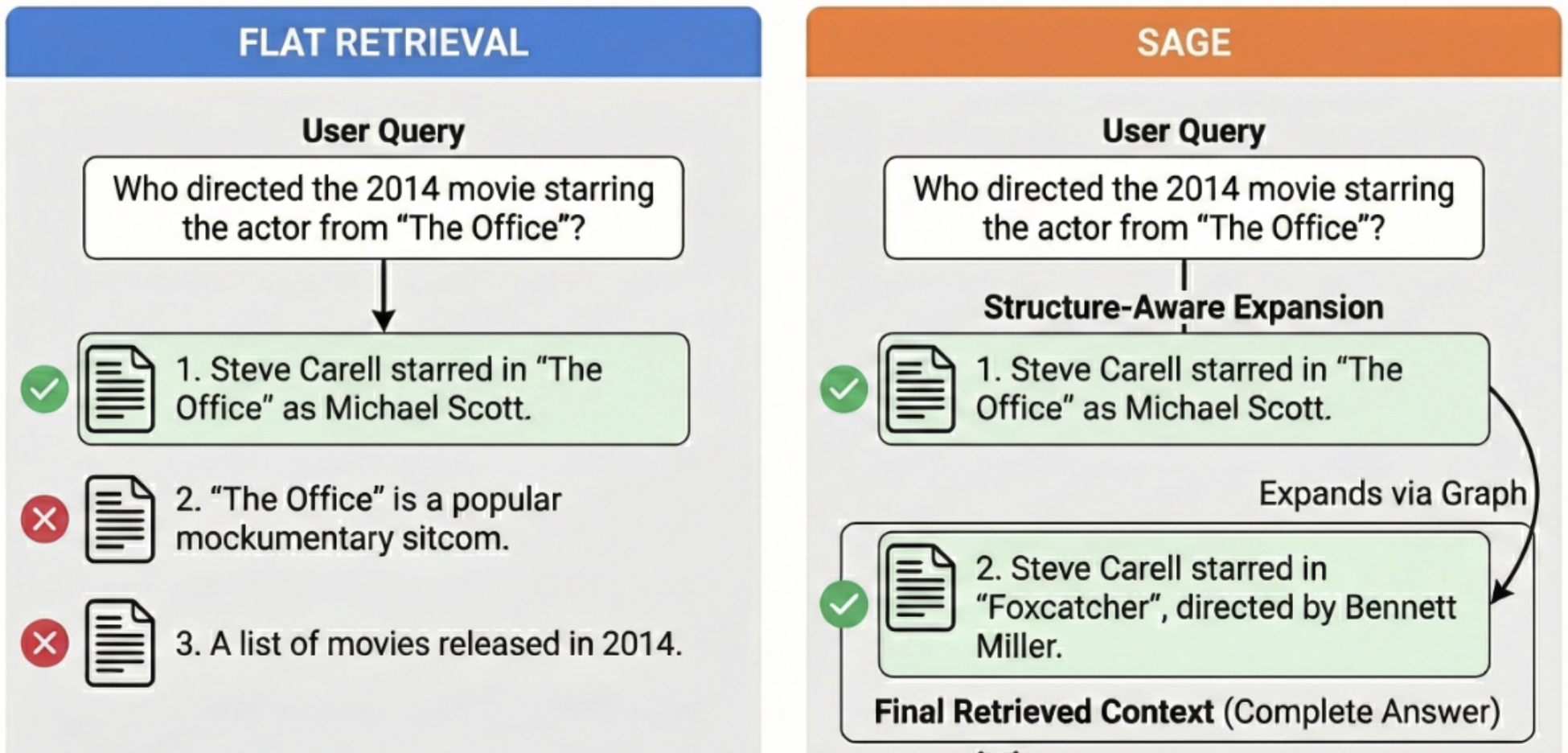}
\caption{\small Illustration of \methodname’s seed$\rightarrow$expand retrieval: starting from an initial relevant chunk, graph expansion retrieves a connected neighbor containing the missing movie/director evidence that flat retrieval fails to surface.}
\label{fig:basic_arch}
\vspace{-1.5em}
\end{figure}

\begin{figure*}[t]
\centering
\includegraphics[width=0.98\textwidth]{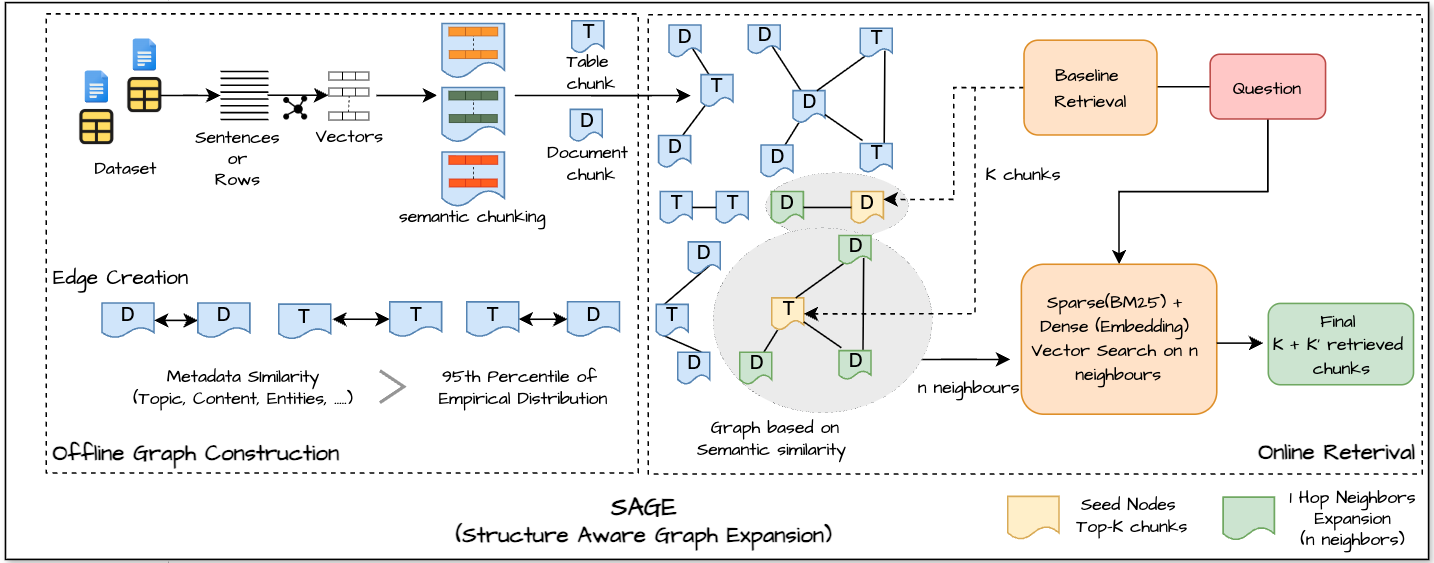}
\vspace{0.25em}
\caption{\small Overview of \methodname. \textbf{Offline:} we semantically chunk documents and tables into nodes and create edges using metadata-driven similarity (e.g., title, topic, content, entities) with percentile-based pruning. \textbf{Online:} a baseline retriever returns $k$ seed nodes, we expand to $n$ first-hop neighbors in the offline graph, and re-rank to select $k'$ additional nodes, yielding a final context of $k{+}k'$.}
\label{fig:main_graph_overview}
\vspace{-0.75em}
\end{figure*}

Flat retrieval treats each chunk independently, ignoring structural dependencies that often connect relevant evidence across documents or modalities. As a result, intermediate evidence that is weakly related to the query in embedding space but structurally connected to relevant content is frequently missed. For example, answering a question about a television show starring Trevor Eyster and its source book requires linking cast information, adaptation metadata, and author details, connections that are unlikely to be recovered through similarity search alone.

To address these limitations, recent work augments RAG with graph-based representations that model relationships across heterogeneous data and enable multi-hop retrieval \cite{edge2024local,jimenez2024hipporag,mavromatis2024gnn}. Prior approaches fall into two categories: \emph{knowledge graphs} (KGs) and \emph{similarity graphs}. KGs encode structured knowledge as typed entity-relation triples and support precise symbolic reasoning via traversal or query languages such as Cypher \cite{Fra18}, and SPARQL \cite{2013sparql}, but require costly entity extraction and schema alignment, making them suitable for domains with stable schemas. Similarity graphs represent text chunks, table segments, or semi-structured nodes with edges induced by embedding similarity, naturally supporting heterogeneous data without explicit entity extraction.

However, their effectiveness depends critically on edge construction: overly dense graphs introduce spurious neighbors, while overly sparse graphs hinder multi-hop reasoning. Node granularity also impacts expressiveness, connectivity, and computational cost. \textbf{Entity-level graphs} capture fine-grained entities, supporting factoid queries but struggling with contextual or aggregate questions.\textbf{ Community-level graphs} group entities or documents into clusters, improving scalability but reducing detail. These trade-offs motivate intermediate representations that balance granularity, scalability, and retrieval effectiveness.

We address this gap by introducing a chunk-level graph that improves recall through structure-aware expansion over semantically coherent units. Rather than retrieving many independent chunks, we first select a small set of seed nodes and expand them via graph traversal, surfacing evidence that may be weakly similar to the query yet strongly connected through document structure, metadata, or shared context. Each node represents a text or table segment, preserving local coherence without requiring explicit entity extraction. The graph is built offline and remains retriever-agnostic, adding minimal runtime overhead. Its compact neighborhood traversal enables efficient multi-hop reasoning and reliable evidence aggregation across heterogeneous corpora. Our contributions are as follows:

\vspace{0.3em}
\noindent\textbullet\ We introduce \methodname, a \textbf{structure-aware retrieval framework} that augments flat retrieval with chunk-level graph expansion, enabling neighbor-aware reasoning while remaining compatible with existing retrievers.\\
\noindent\textbullet\ Through \textbf{similarity-induced graph expansion}, \methodname consistently improves recall over strong flat hybrid baselines under fixed budgets (+5.7 on OTT-QA, +8.5 on STaRK), with the largest gains on implicit multi-hop and cross-modal queries.\\
\noindent\textbullet\ For explicit schema graphs, we further propose \agentname, a retrieval agent that performs \textbf{schema-constrained expansion}, enforcing structural validity during traversal and outperforming naive dense expansion.

\section{\methodname Approach}
\label{sec:method}

Given a corpus represented as a graph $G=(V,E)$ with node content $\mathrm{text}(v)$ and optional metadata $\mathrm{meta}(v)$, the retrieval goal for a query $q$ is to return a ranked list of evidence nodes $R_k(q)$ such that relevant evidence appears in the top-$k$ positions.

\subsection{Offline Graph Construction}
\label{sec:offline-graph}
We construct heterogeneous graphs $G$ over document chunks and table segments (for document graphs) or preserve native entity-relation schemas (for KGs). The graph construction process is dataset-agnostic and occurs offline before query time.
\paragraph{A. Data Processing and  Node Creation}
\paragraph{1. Semantic chunking of documents.}
We segment each document into a sequence of sentences and embed overlapping context windows $w_i = [s_{i-1}, s_i, s_{i+1}]$ (with boundary truncation) using a sentence encoder \cite{reimers2019sentence}. Let $\mathbf{h}_i$ denote the embedding of $w_i$. We compute cosine similarity between adjacent windows as $s_i = \cos(\mathbf{h}_i, \mathbf{h}_{i+1})$.

We introduce a boundary at indices where local coherence drops, using a percentile-based rule over $\{s_i\}$: letting $\theta = \mathrm{Percentile}_{20}(\{s_i\})$ (equivalently, selecting the top-80\% ``drop'' positions), we split at all $i$ such that $s_i \le \theta$.
After chunking, we use an LLM to generate per-chunk metadata (topic, title, and entities) for downstream graph construction.

\paragraph{2. Table segmentation.}
We use an LLM to generate a short table description and column descriptions; these become part of the table-chunk metadata. For each table, we retain table title/description, column headers, and column descriptions, and 5--10 rows per segment to preserve vertical context while keeping segments within a bounded context size.

\paragraph{3. Semi-structured nodes.}
For semi-structured nodes, we treat each object as a ``chunk'' and construct metadata from its textual fields (e.g., name/title and long-form descriptions). These metadata fields are used to build downstream graph connectivity. 

\paragraph{B. Edge creation}
\paragraph{Graph construction via metadata similarities.}
We embed these metadata fields using a Sentence-Transformer encoder and connect nodes based on modality-specific similarity signals: (1) \textbf{Document--Document} edges capture topic-topic similarity, content-content similarity, and entity overlap; (2) \textbf{Table--Table} edges capture column-column similarity, title-title similarity, and entity overlap; (3) \textbf{Table--Document} edges capture content-column similarity, topic-title similarity, and entity overlap; (4) \textbf{Semi-Structured Chunk} edges connect objects via similarity over their descriptive metadata.

We only keep a similarity edge when its metadata similarity exceeds the 95th percentile of the empirical distribution, which maintains sparsity and reduces dense-neighborhood noise. We also preserve parent/structural links (e.g., chunks from the same source) and retain explicit schema edges when available. Figure~\ref{fig:main_graph_overview} illustrates the resulting graph structure.

\paragraph{Edge metadata for traversal.}
Edges store lightweight metadata to guide controlled graph traversal. Table--table edges record joinable column names, document--document edges record shared entities and confidence scores, and table--document edges record shared entities and, when available, row or column references. This metadata allows multi-hop reasoning to prioritize neighbors that align with query entities or satisfy structural constraints.

\subsection{Online Retrieval}

At query time, we employ a two-stage retrieval strategy: (i) \textbf{initial baseline retrieval} to obtain $k$ seed nodes, and (ii) \textbf{graph-based neighbor expansion and pruning} to select $k'$ additional nodes. The baseline retrieval method differs between document graphs and KGs due to their distinct structural properties: hybrid sparse-dense retrieval works well for OTT-QA's text/table data, while \agentname is necessary for STaRK's schema-rich KGs (as hybrid retrieval performs poorly on STaRK, shown in Table~\ref{tab:graph_deltas_qcol}).

\paragraph{A. Initial baseline retrieval.}

\paragraph{1. Similarity graphs}
For document graphs (tables and text chunks), we use a hybrid sparse-dense retrieval baseline. We compute BM25 \cite{robertson2009probabilistic} scores for lexical matching and cosine similarity over dense embeddings for semantic matching, then combine these scores to select the top-$k$ seed nodes. This approach works well when chunks contain sufficient textual content and when relevance can be determined from surface-level semantic similarity.

\paragraph{2. Semi-Structured Knowledge Bases (SKBs)} For semi-structured knowledge bases with explicit schemas (i.e., graphs whose nodes are associated with textual fields), BM25 and cosine similarity are limited. They ignore edge types and typed relations, cannot reason over multi-hop paths, and exhibit type bias, often retrieving nodes of the same type rather than the correct entities required by the query.

\begin{figure}[h]
\vspace{-0.75em}
    \centering
\includegraphics[width=0.98\linewidth]{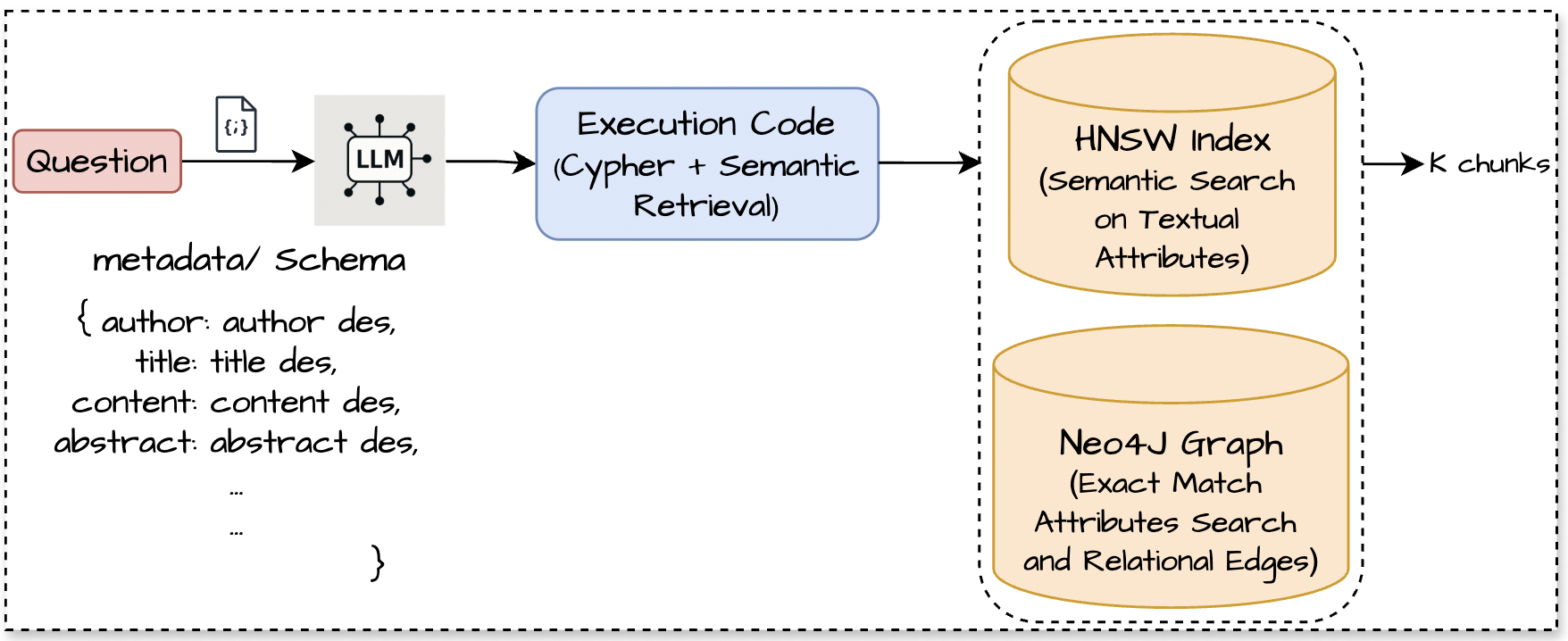}
\vspace{0.25em}
\caption{\small SKB baseline retrieval: the question and schema metadata guide an LLM agent to generate a retrieval plan interleaving HNSW semantic search with Cypher symbolic queries.}
\label{fig:stark_retrieval}
\vspace{-0.75em}
\end{figure}

To address this, we introduce \agentname, an LLM-based agent for SKB retrieval. Given a question, the agent has access to the entity schemas, relations, and two retrieval tools. It generates a multi-step plan that alternates between: \textbf{(1) Semantic Candidate Generation}: retrieving nodes via Hierarchical Navigable Small World (HNSW) \cite{malkov2018efficient} search based on textual similarity and descriptive attributes. This enables flexible matching, allowing the agent to identify semantically relevant nodes even when query keywords do not appear explicitly. \textbf{(2) Structural Expansion}: traversing edges with Cypher queries in Neo4j \cite{neo4j2026}, constrained by entity types and relations to ensure multi-hop structural correctness. This guarantees that retrieved nodes satisfy explicit relational and type constraints, which simple keyword or similarity-based methods cannot enforce.

We execute the generated plan and return the top-$k$ retrieved nodes as seeds. A fallback to BM25 and cosine similarity is used when the plan returns no results to ensure robustness, as the agent may fail to generate correct multi-step queries for certain questions or encounter nodes with sparse textual content. This combination balances the precision of schema-aware traversal with the coverage of traditional retrieval methods.

\paragraph{B. Graph-based neighbor expansion and pruning.}
Regardless of the baseline, we apply a unified graph expansion strategy to both documents and KGs. \agentname is simply one possible seed retriever specialized for schema-rich KGs; the \methodname expansion operator itself remains invariant across all datasets and seed retrievers. Given the $k$ seed nodes from initial retrieval: (1) \textbf{Neighbor Expansion:} collect all first-hop neighbors of the $k$ seeds from the pre-built graph $G$, producing $n$ candidate nodes; (2) \textbf{Neighbor Pruning:} re-rank the $n$ candidates using BM25 and dense embedding similarity with respect to the query, and select the top-$k'$ neighbors; (3) \textbf{Final Context:} return the union of the original $k$ seeds and the $k'$ neighbors, yielding $k{+}k'$ total nodes.

{
\vspace{-1.0em}
\begin{lstlisting}[caption={Unified $k{+}k'$ retrieval (pseudo-code)}]
Input: query q, budgets k,k', baseline retriever BL, graph G
S = TopK(BL(q), k)
N = OneHopNeighbors(G, S) 
R = TopK(BM25DenseRank(q, N), k') 
Return Union(S, R) 
\end{lstlisting}
\vspace{-0.5em}
}
This approach is more efficient than computing full node-to-node similarity at runtime. Pre-computed connections capture bridging evidence, reduce noise by focusing on relevant neighbors, preserve multi-hop context, and scale to large graphs without exhaustive pairwise comparisons.
\section{Experiments}
\label{sec:experiments}

\paragraph{Datasets.}
We evaluate on OTT-QA and STaRK~\cite{wu24}. OTT-QA is derived from Wikipedia and includes both text passages and tables, where each table is originally linked to a set of documents. We use the dev split to construct our graph and rebuild all connections from scratch instead of using the predefined links.

STaRK contains semi-structured knowledge graphs across three domains: AMAZON (product recommendation with limited node types but rich textual content), MAG (academic citations over papers and authors with one- and two-hop reasoning), and PRIME (biomedical knowledge with dense relational structure where all nodes are answerable). We add similarity edges only among nodes with rich textual descriptions(e.g., products, papers, and biomedical entities) while atomic entities with minimal text are handled via lexical Cypher matching. Table~\ref{tab:dataset_overview} summarizes the resulting graph statistics.

Each STaRK subset provides synthetic and human evaluation splits. Synthetic queries are generated through a four-step pipeline that combines relational templates with LLM-extracted properties to produce diverse multi-hop questions with verified ground truth, while human queries reflect realistic styles, ambiguity, and linguistic variation.

\begin{table}[h]
\vspace{-0.75em}
\centering
\setlength{\aboverulesep}{0.25pt}
\setlength{\belowrulesep}{0.25pt}
\scriptsize
\setlength{\tabcolsep}{0.5pt}
\begin{tabular}{lcccc}
\toprule
\bf Dataset &\bf \#Nodes &\bf \#Node types &\bf \#Edges & \bf \#Edge Types \\
\midrule
\textbf{OTT-QA} & 22,886 & 2 & 751,836 & 3 \\
\hline\noalign{\vskip 1pt}
\multicolumn{5}{c}{\textbf{STaRK}} \\
\hline\noalign{\vskip 2pt}
\bf AMAZON & 13,920,204 & 2 & 40,876,246 & 4 \\
\bf MAG    & 1,872,968 & 4 & 36,853,636 & 5 \\
\bf PRIME  & 129,375 & 10 & 9,825,282 & 19 \\
\bottomrule
\end{tabular}
\vspace{0.5em}
\caption{\small Overview of datasets used in our experiments. Node and edge counts are total numbers in the graph. Node and edge types indicate schema diversity.}
\vspace{-0.75em}
\label{tab:dataset_overview}
\end{table}

These datasets represent two complementary regimes of heterogeneous data. In our OTT-QA setting, structure is implicit: links between text and table chunks are induced from content similarity and metadata overlap rather than predefined relations. In contrast, STaRK provides explicit structure, where typed entities and relations follow a fixed schema. This contrast lets us evaluate whether \methodname adapts its retrieval operators to the underlying topology

Further details on node and edge types are in Appendix~\ref{sec:dataset_stats} .

\paragraph{Baselines for OTT-QA}
Prior OTT-QA work mainly targets table-level retrieval, as in CRAFT \cite{singh2025craft} and OTTeR \cite{huang-etal-2022-mixed}, rather than chunk-level retrieval. Many methods also depend on predefined table-document links, including COS \cite{ma-etal-2023-chain}, RINK \cite{Park2023RINKRE}, CARP \cite{zhong2022reasoning}, and Dual Reader-Parser \cite{chen2020open}. Methods that ignore these links use different setups: ARM \cite{chen-etal-2025-retrieve} adopts gold-set evaluation, while CORE \cite{ma-etal-2022-open-domain} attempts to reconstruct the links, preventing direct comparison. Extended Baseline (BL) retrieves the same number of chunks directly without graph-based expansion.

\vspace{-0.2em}
\paragraph{Baselines for STaRK}
For STaRK, we compare against dense retrievers (\textbf{ada-002}, \textbf{multi-ada-002}, \textbf{DPR} \cite{karpukhin2020dense}), a sparse lexical baseline (\textbf{BM25}), and graph-based reasoning with \textbf{QA-GNN} \cite{yasunaga2022qagnnreasoninglanguagemodels}. Hybrid retrieval methods, including \textbf{4StepFocus} \cite{Boe24} and \textbf{FocusedR} \cite{boer2025focus}, combine vector search with LLM-based triplet extraction and iterative refinement. Query expansion approaches (\textbf{HyDE} \cite{gao2022hyde}, \textbf{AGR} \cite{chen-etal-2024-analyze}, \textbf{RAR} \cite{shen-etal-2024-retrieval}, \textbf{KAR} \cite{Xia24}) reformulate queries using KG or LLM-derived signals. Agentic systems include \textbf{ReACT} \cite{yao2023reactsynergizingreasoningacting}, which interleaves reasoning and retrieval, and \textbf{Reflexion} \cite{shinn2023reflexion}, which adds episodic memory for iterative improvement. Fine-tuned semi-structured retrievers such as \textbf{mFAR} \cite{Li24} and \textbf{MoR} \cite{Lei25} adapt field weighting or planning with graph traversal, while \textbf{AvaTaR} \cite{wu2024avatar} optimizes tool-using agents via contrastive comparator-based refinement and a memory of past failures, which we categorize as a fine-tuned approach.

\vspace{-0.5em}
\paragraph{Evaluation Metrics.} For \textbf{OTT-QA}, retrieval is measured using Recall@$k$, which assesses how many relevant passages are retrieved. While we focus only on recall for \textbf{STaRK}, we report standard metrics from the original paper: Hit@1, Hit@5, Recall@20, and MRR, capturing both the ranking quality and coverage of retrieved knowledge.

\subsection{OTT-QA: Hybrid Text-Table Results}
\label{sec:ott_results}

\begin{figure}[t]
\centering
\includegraphics[width=0.98\linewidth]{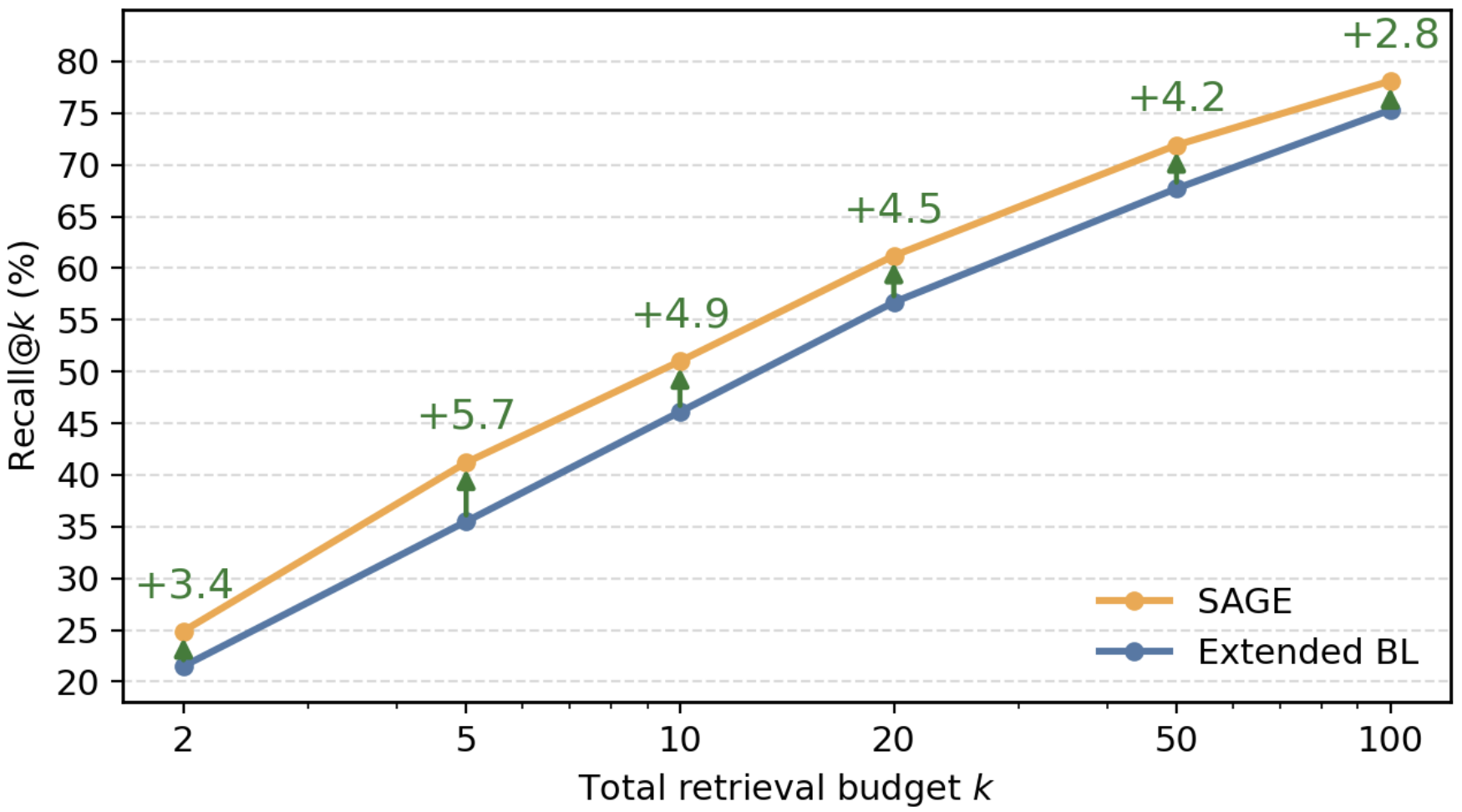}
\caption{\small OTT-QA retrieval Recall@$k$ (\%, higher is better). BL (Baseline) is a flat hybrid retriever combining BM25 and dense embeddings. BL+Graph retrieves $k_1$ seeds with BL and adds $k_2$ one-hop neighbors from the induced graph ($k=k_1+k_2$), while Extended BL retrieves $k$ items directly with BL. Arrows annotate the absolute gain at each $k$.}
\vspace{-1.5em}
\label{fig:ottqa_recall}
\end{figure}

\paragraph{Graph-based retrieval improves Recall@$k$ across all retrieval sizes.}
Graph-based retrieval consistently improves Recall@$k$ across all retrieval sizes (Figure~\ref{fig:ottqa_recall}). By adding one-hop neighbors from the chunk-level graph, the number of relevant items retrieved increases at every $k$, from very small retrieval sets ($k \approx 2$) to larger sets ($k=100$). This demonstrates that the graph effectively surfaces additional relevant nodes that flat similarity-based retrieval alone cannot capture.

\paragraph{Graph impact is larger for smaller retrieval sets.}
The impact of the graph is particularly pronounced for smaller retrieval sets. For instance, recall increases by 5.7 points at $k=20$ compared to 2.8 points at $k=100$, showing that relative improvements are largest when fewer items are retrieved. This highlights that structural connections in the graph are most beneficial when retrieval is constrained, allowing the system to efficiently identify relevant neighbors that would otherwise be missed.

\begin{table*}[t]
\setlength{\aboverulesep}{0.1pt}
\setlength{\belowrulesep}{0.1pt}
\small
\centering
\setlength{\tabcolsep}{5pt}
\begin{tabular}{l l|cc|cc|cc}
\toprule
\bf Category & \bf Method & \multicolumn{2}{c|}{\bf AMAZON} & \multicolumn{2}{c|}{\bf MAG} & \multicolumn{2}{c}{\bf PRIME} \\
 &  & \bf Human & \bf Synthetic & \bf Human & \bf Synthetic & \bf Human & \bf Synthetic \\
 &  & $n=\,$81 & $n=\,$1638 & $n=\,$84 & $n=\,$2664 & $n=\,$98 & $n=\,$2801 \\
\midrule

\multirow{3}{*}{Dense}
 & ada-002 & 35.46 & 53.29 & 35.95 & 48.36 & 41.09 & 36.00 \\
 & multi-ada-002 & 40.22 & 55.12 & 39.85 & 50.80 & 47.21 & 38.05 \\
 & DPR & 15.23 & 44.49 & 25.00 & 42.11 & 10.69 & 30.13 \\
\cmidrule(lr){1-8}

\multirow{1}{*}{Sparse}
 & BM25 & 15.23 & 53.77 & 32.46 & 45.69 & 42.32 & 31.25 \\
\cmidrule(lr){1-8}

\multirow{1}{*}{Structural}
 & QAGNN & 21.54 & 52.05 & 28.76 & 46.97 & 17.62 & 29.63 \\
\cmidrule(lr){1-8}

\multirow{3}{*}{Hybrid}
 & 4StepFocus & -- & 56.20 & -- & 65.80 & -- & 55.90 \\
 & FocusedR. & -- & 56.00 & -- & \textbf{78.80} & -- & \textbf{65.50} \\
 & Dense+BM25 & 24.30 & 51.49 & \underline{46.64} & 51.49 & 52.65 & 39.11 \\
\cmidrule(lr){1-8}

\multirow{4}{*}{Query Expansion}
 & HyDe & 39.25 & 53.71 & 37.23 & 50.02 & 47.70 & 43.55 \\
 & RAR & 36.15 & 54.63 & 35.19 & 50.87 & 49.01 & 44.50 \\
 & AGR & 37.27 & 53.38 & 37.23 & 51.89 & 49.65 & 46.63 \\
 & KAR & \underline{40.62} & \underline{57.29} & 46.60 & 60.28 & \underline{59.90} & 50.81 \\
\cmidrule(lr){1-8}

\multirow{2}{*}{Agentic}
 & ReACT & 35.95 & 50.81 & 35.95 & 47.03 & 41.09 & 33.63 \\
 & Reflexion & 35.95 & 54.70 & 35.95 & 49.55 & 41.09 & 38.52 \\
\cmidrule(lr){1-8}

\multirow{3}{*}{Finetuned}
 & AvaTaR & 42.43 & 60.57 & 35.94 & 49.70 & 53.34 & 42.23 \\
 & mFAR & -- & 58.50 & -- & 71.70 & -- & 68.30 \\
 & MoR & -- & 59.90 & -- & 75.00 & -- & 63.50 \\
\cmidrule(lr){1-8}

\multirow{3}{*}{SAGE (Ours)}
 & \agentname & 33.56 & 53.95 & 53.51 & 66.40 & 63.56 & 57.64 \\
 & \agentname+\methodname+Edge & 36.93 & 54.12 & 55.90 & 68.60 & 65.85 & 58.42 \\
 & \agentname+\methodname-Edge & \textbf{42.05} & \textbf{58.28} & \textbf{56.40} & \underline{70.10} & \textbf{66.37} & \underline{60.98} \\
\bottomrule
\end{tabular}
\vspace{0.25em}
\caption{\small Recall@20 performance on Human and Synthetic queries across all STaRK datasets (AMAZON, MAG, PRIME). Highest non-finetuned method is bolded, second highest non-finetuned method is underlined. Note: FocusedR, 4StepFocus, mFAR, and MoR report results only on the Synthetic split; Human split results are not available from their publications or released artifacts.}

\label{tab:r20_combined}
\end{table*}

\paragraph{Effect of metadata on graph building}
We analyze metadata similarities to guide graph construction. Table--table pairs show strong correlation between title and description similarity (0.846, Table~\ref{tab:tab_tab_matrix}), while document--document pairs align across topic and content similarity (0.650, Table~\ref{tab:doc_doc_matrix}). Cross-modal document--table correlations range from 0.517 to 0.691 (Table~\ref{tab:doc_tab_matrix}). These patterns indicate that similarity along one metadata dimension often predicts similarity along others, helping form coherent graph neighborhoods.

Similarity distributions are largely bell-shaped with a high-end tail, providing broad coverage while emphasizing strongly related chunks. Because most chunks contain a single entity, entity overlap becomes a key bridge: chunks sharing an entity can connect otherwise distinct topics. Percentile-based pruning that retains the top 5\% of similarities preserves these meaningful links without over-densifying the graph. Additional distributions and correlation analyses appear in Appendix~\ref{sec:ott_graph_building_stats}.

\begin{table}[t]
\scriptsize
\setlength{\aboverulesep}{0.1pt}
\setlength{\belowrulesep}{0.1pt}
\centering
\setlength{\tabcolsep}{3.0pt}
\begin{tabular}{lccc}
\toprule
\bf Similarity metric & \bf doc--doc & \bf doc--table & \bf table--table \\
\midrule
topic\_similarity & 0.350 & 0.247 & 0.263 \\
content\_similarity & 0.389 & 0.350 & 0.404 \\
column\_similarity & 0.000 & 0.242 & 0.409 \\
entity\_relationship\_overlap & 0.013 & 0.000 & 0.000 \\
entity\_count & 0.682 & 0.477 & 0.849 \\
topic\_title\_similarity & 0.000 & 0.221 & 0.000 \\
topic\_summary\_similarity & 0.000 & 0.236 & 0.000 \\
title\_similarity & 0.000 & 0.000 & 0.239 \\
description\_similarity & 0.000 & 0.000 & 0.262 \\
\bottomrule
\end{tabular}
\vspace{0.5em}
\caption{\small Cross-modal similarity statistics across document and table pairs in OTT-QA.}
\vspace{-1.5em}
\label{tab:cross_modal_similarity}
\end{table}

Table~\ref{tab:cross_modal_similarity} further reveals three cross-modal structural effects.
\textbf{Entity-centric connectivity is dominant but modality dependent.}
Shared entities form the backbone of the graph, but their influence varies by node type. Tables create dense entity hubs, while text regions support more semantically blended neighborhoods, yielding a hybrid topology that enables both factual linking and thematic traversal.
\textbf{Modalities encode fundamentally different structural signals.}
Text captures narrative and relational structure, while tables provide schema-level alignment unavailable in prose. Similarity is therefore asymmetric, and effective graph construction must treat modalities as complementary rather than interchangeable.
\textbf{Cross-modal alignment emerges from signal composition rather than dominance.}
Document-table links arise from aggregating multiple weak but consistent signals, not a single dominant metric. Reliable cross-modal bridges require multi-signal fusion, where stability comes from agreement across dimensions.

\subsection{STaRK: Semi-Structured Data Results}
\label{sec:stark_results}

\paragraph{Effect of graph on agentic retriever}
Similar to our observations on OTT-QA, graph-based retrieval also improves Recall@20 for agentic retrievers across all STaRK datasets (Table~\ref{tab:r20_combined}). By expanding the retrieved set to include connected nodes, the graph surfaces additional relevant nodes that the base agent misses, leading to higher recall on both Human and Synthetic queries.

\paragraph{Are there dataset-specific patterns where graph augmentation helps more?} Graph augmentation yields larger gains when nodes contain richer textual content. We evaluate two expansion policies. \agentname+\methodname-Edge skips neighbor expansion when Cypher queries include explicit edges, preserving the semantic relationships specified in the query. In contrast, \agentname+\methodname+Edge always expands neighbors. On AMAZON, where product metadata provides detailed node descriptions, \agentname+\methodname-Edge improves R@20 by +8.49 on Human queries and +4.33 on Synthetic queries. Datasets with sparser node information, such as MAG and PRIME, show smaller improvements of roughly 1 to 2 points. This pattern suggests that chunk-level graphs are most effective when nodes contain enough context to support meaningful multi-hop evidence chains, enabling the retrieval of connections that flat retrieval methods fail to capture.

\paragraph{How does this baseline+graph compare vs other state of the art methods}
\agentname+\methodname-Edge substantially outperforms traditional baselines, including Dense, Sparse, and Query Expansion methods, across all datasets (Table~\ref{tab:r20_combined}), while matching the performance of finetuned approaches such as AvaTaR, mFAR, and MoR. FocusedR achieves higher scores on MAG and PRIME, where structured KG triples provide a strong signal, but \agentname+\methodname-Edge performs best on AMAZON by leveraging rich node content for chunk-level expansion. Importantly, \agentname+\methodname-Edge requires only a single LLM call and one retrieval step, making it significantly faster and more practical for large-scale deployment than multi-iteration alternatives.

\paragraph{How efficient is our Agentic approach vs other state of the art methods}\begin{table}[t]
\small
\setlength{\aboverulesep}{0.1pt}
\setlength{\belowrulesep}{0.1pt}
\centering
\setlength{\tabcolsep}{3.0pt}
\begin{tabular}{lcccc}
\toprule
\bf Method &\bf  \#Retr. &\bf  \#LLM &\bf  \#Iter. &\bf  Set Join \\
\midrule
4StepFocus &
1 + n &
2 &
n &
Triplet cand. \\

FocusedR &
$\ge$3+n &
3 &
$n_1 \!\times\! n_2$ &
Sym. cand \\

AvaTaR &
n &
n+1 &
n &
--  \\

KAR &
e+3&
2 &
\textbf{1} &
--  \\

ReACT &
n &
n+1 &
n &
--\\

Reflexion &
n  &
(n+1)E+R &
nE+R &
-- \\

\textbf{SAGE} &
\textbf{2} &
\textbf{1} &
\textbf{1} &
--\\
\bottomrule
\end{tabular}
\vspace{0.5em}
\caption{\small 
Theoretical runtime decomposition from method descriptions.
\#Retr. counts number of retrieval operations 
$n$ is a method-dependent number of reasoning steps.
$n_1, n_2$ denote outer/inner iterations.
$e$ is the number of extracted entities.
$E$ and $R$ denote episodes and calls. Set Join indicates explicit candidate-neighborhood intersections.
}
\vspace{-1.5em}
\label{tab:theoretical_runtime}
\end{table}
Table~\ref{tab:theoretical_runtime} exposes three major efficiency bottlenecks common to existing retrieval methods. First, repeated LLM invocations substantially increase inference cost. Second, iterative retrieval loops introduce execution paths that can grow arbitrarily long. Third, symbolic set-join operations add significant computational overhead. ReAct, Reflexion, and AvaTaR rely on step-by-step retrieval, issuing an LLM call at each action step, which sharply increases both latency and cost as the number of steps grows. In contrast, 4StepFocus and FocusedR perform repeated symbolic pruning with set intersections across multiple passes, leading to substantial runtime growth as query complexity increases. KAR reduces LLM usage to two calls, but still requires multiple dataset-wide retrievals, which limits scalability.

\agentname eliminates all three bottlenecks. It operates with a single LLM call, a single retrieval, and one graph expansion. There are no iterative loops and no symbolic set-join operations. This produces a fixed, lightweight execution path that is dramatically faster while maintaining competitive recall. For real-world deployments where latency and cost are paramount, \agentname delivers a superior accuracy-per-compute tradeoff. Details about order complexity calculation is in Appendix~\ref{sec:runtime}.

\subsection{Ablation Study}

\paragraph{Sensitivity Analysis: Effect of percentile during edge creation} We study how percentile-based pruning shapes graph connectivity in the OTT-QA setting. Pruning controls the trade-off between connectivity and noise by limiting similarity edges per chunk.

We compare two edge selection strategies. \textbf{OR logic} retains edges satisfying any similarity criterion, encouraging connectivity across heterogeneous signals, while \textbf{AND logic} requires all criteria and heavily prunes the graph.

\begin{table}[t]
\scriptsize
\setlength{\aboverulesep}{0.1pt}
\setlength{\belowrulesep}{0.1pt}
\centering
\setlength{\tabcolsep}{3.0pt}
\begin{tabular}{cccccc}
\toprule
\bf Percentile & \bf Logic & \bf Total Edges & \bf \%Cand. & \bf Avg Deg. & \bf Density \\
\midrule
80 & OR  & 1,240,504 & 39.5\% & 109.2 & 0.0048 \\
85 & OR  &   969,911 & 30.9\% & 85.4  & 0.0038 \\
90 & OR  &   669,992 & 21.3\% & 59.0  & 0.0026 \\
\textbf{95} & \textbf{OR}  &   \textbf{344,955} & \textbf{11.0\%} & \textbf{30.4}  & \textbf{0.0013} \\
97 & OR  &   206,461 & 6.6\%  & 18.2  & 0.0008 \\
99 & OR  &    60,371 & 1.9\%  & 5.3   & 0.0002 \\
\midrule
80 & AND &   150,973 & 4.8\%  & 13.3  & 0.0006 \\
85 & AND &   101,963 & 3.2\%  & 9.0   & 0.0004 \\
90 & AND &    63,042 & 2.0\%  & 5.6   & 0.0002 \\
95 & AND &    30,920 & 1.0\%  & 2.7   & 0.0001 \\
97 & AND &    21,553 & 0.7\%  & 1.9   & 0.0001 \\
99 & AND &    13,434 & 0.4\%  & 1.2   & 0.0001 \\
\bottomrule
\end{tabular}
\vspace{0.5em}
\caption{\small
Impact of pruning percentile and logical edge selection on graph sparsity.
}
\vspace{-1.5em}
\label{tab:pruning_impact}
\end{table}

Table~\ref{tab:pruning_impact} shows that AND logic consistently produces extremely low
average degree (1.2--2.7), yielding weak connectivity and fragmented components. This prevents reliable multi-hop traversal and offers no practical benefit. We therefore adopt OR logic.

Under OR logic, the 95th percentile offers the best balance. Stricter thresholds fragment neighborhoods and reduce expansion coverage, whereas looser pruning sharply increases degree and noise. The 95th percentile maintains stable connectivity while keeping candidate expansion and reranking tractable.

The 95th percentile targets approximately 5--25 effective neighbors per chunk after expansion and downstream filtering, preserving connectivity while keeping reranking tractable.

\paragraph{How much of the performance gain is due to the graph structure itself versus the graph expansion strategy?}

Both variants outperform the base \agentname model (Table~\ref{tab:r20_combined}), but \agentname+\methodname-Edge consistently delivers larger gains. For example, on AMAZON Human queries it improves R@20 by +8.49 compared to +6.21 for \agentname+\methodname+Edge, and on MAG Human queries it adds +1.31 versus +0.84. This trend holds across datasets. Selectively skipping expansion for edge-aware queries prevents irrelevant neighbors from diluting semantically grounded candidates, while expansion remains beneficial for queries without structural constraints. These results indicate that improvements stem from the interaction between the graph and the expansion policy: the graph supplies useful connectivity, and conditioning expansion on query semantics ensures that only meaningful neighbors are introduced.

\paragraph{What is the effect of the initial retriever on the graph-based expansion?}

\begin{table}[t]
\setlength{\aboverulesep}{0.25pt}
\setlength{\belowrulesep}{0.25pt}
\scriptsize
\centering
\setlength{\tabcolsep}{3.5pt}
\begin{tabular}{l|ccc|ccc}
\toprule
& \multicolumn{3}{c|}{\textbf{Human}} & \multicolumn{3}{c}{\textbf{Synthetic}} \\
\cline{2-7}
\noalign{\vskip 2pt}
\textbf{Method} & A. & M. & P. & A. & M. & P. \\
\midrule
Hybrid & 24.3 & 46.6 & 52.7 & 51.5 & 51.5 & 39.1 \\
Hybrid+Graph & 31.1 & 46.4 & 53.4 & 53.8 & 52.9 & 38.4 \\
\cmidrule(lr){2-7}
\textbf{Change ($\Delta$)} & 
\textcolor{DarkGreen}{+6.8} & \textcolor{DarkRed}{-0.2} & \textcolor{DarkGreen}{+0.7} & 
\textcolor{DarkGreen}{+2.3} & \textcolor{DarkGreen}{+1.4} & \textcolor{DarkRed}{-0.7} \\
\midrule
\agentname & 33.6 & 53.5 & 63.6 & 54.1 & 66.4 & 58.4 \\
\agentname+\methodname & 42.1 & 56.4 & 66.4 & 58.3 & 70.1 & 61.0 \\
\cmidrule(lr){2-7}
\textbf{Change ($\Delta$)} & 
\textcolor{DarkGreen}{\textbf{+8.5}} & \textcolor{DarkGreen}{\textbf{+2.9}} & \textcolor{DarkGreen}{\textbf{+2.8}} & 
\textcolor{DarkGreen}{\textbf{+4.2}} & \textcolor{DarkGreen}{\textbf{+3.7}} & \textcolor{DarkGreen}{\textbf{+2.6}} \\
\bottomrule
\end{tabular}
\vspace{0.25em}
\caption{\small Recall@20 before and after graph augmentation across different baseline retrievers on STaRK datasets. Hybrid refers to BM25+Cosine similarity. A., M., and P. stand for AMAZON, MAG, and PRIME, respectively..}
\label{tab:graph_deltas_qcol}
\vspace{-1.0em}
\end{table}

The effectiveness of graph augmentation depends strongly on the quality of the initial retriever, as shown in Table~\ref{tab:graph_deltas_qcol}. For the dense and sparse hybrid baseline, Recall@20 is substantially lower than that of the agent retriever, and the improvement from graph expansion is limited. In some cases, particularly on the PRIME and MAG subsets at low retrieval depths, the delta is negligible or even negative, for example, Hybrid Human MAG goes from 46.6 to 46.4. This is partly because the hybrid retriever does not take node type into account, while graph expansion tends to favor nodes with embeddings similar to the seed nodes, which are often of the same type. Without a strong initial candidate set and guidance on node type, the expansion can introduce irrelevant nodes, limiting gains.

By contrast, the agent retriever provides a higher initial Recall@20 and determines the node type. Graph expansion with the agent consistently improves performance across all datasets and query types, for example, AMAZON Human goes from 33.6 to 42.1. The combination of stronger initial candidates and type-aware retrieval allows the graph to add relevant nodes, amplifying the benefits of expansion. Full Recall@k curves showing these trends are provided in Appendix~\ref{sec:stark_k}.

\subsection{Error Analysis on STaRK Human}
We manually analyze retrieval failures on STaRK Human set and categorize errors into three classes:

\textbf{- Data Errors} reflect limitations in the dataset itself: These contain: \textbf{(1) Structural}, where there are missing or misdirected edges in the KG.\textbf{(2) Under-Specification}, where queries lack enough constraints to identify the correct entity.
\textbf{(3) Evaluation Artifacts}, where gold annotations are incomplete, marking only a subset of valid answers. 

\textbf{- Cypher Generation Errors} arise when translating natural language to graph queries: \textbf{(1) Constraint}, where numeric, temporal, or logical filters are not correctly applied; and \textbf{(2) Structure}, where the system fails to select the correct node or edge types for graph traversal.

\textbf{- Runtime Errors} occur during query execution despite syntactically correct Cypher: \textbf{(1) Semantic}, where mismatches in units, terminology, or abbreviations prevent correct retrieval; and \textbf{(2) Over-Retrieval}, where queries return excessive candidates due to insufficient filtering.

\begin{table}[h]
\vspace{-0.5em}
\scriptsize
\setlength{\aboverulesep}{0.25pt}
\setlength{\belowrulesep}{0.25pt}
\centering
\setlength{\tabcolsep}{4pt}
\begin{tabular}{l|l|ccc}
\toprule
\textbf{Category} & \textbf{Method} &\bf  Amazon & \bf MAG &\bf  Prime \\
\midrule
\multirow{3}{*}{Data} 
& Structural & 5 & 6 & 4 \\
& Under-Spec & 4 & 5 & 3 \\
& Eval Artifacts & 7 & 4 & 4 \\
\midrule
\multirow{2}{*}{Cypher}
& Constraint & 36 & 30 & 31 \\
& Structure & 0 & 0 & 22 \\
\midrule
\multirow{2}{*}{Runtime}
& Semantic & 25 & 37 & 12 \\
& Over-Retrieval & 23 & 18 & 22 \\
\midrule
& \textbf{Total} & 100 & 100 & 100 \\
\bottomrule
\end{tabular}
\vspace{0.25em}
\caption{\small Error distribution (\%) across STaRK datasets.}
\label{tab:error_multirow}
\vspace{-1.0em}
\end{table}

Constraint extraction failures dominate across AMAZON and PRIME, reflecting challenges in translating complex product attributes or biomedical predicates into precise graph queries. MAG is primarily affected by semantic mismatches, due to dense technical abstracts where terminology and phrasing differ between queries and graph entities. Over-retrieval occurs in all datasets when filters are insufficient. PRIME uniquely exhibits substantial graph structure errors, consistent with its richer schema requiring correct edge and node selection. Data errors are generally minor but still contribute small percentages, particularly in MAG where gold annotations are incomplete (Table~\ref{tab:error_multirow}).

Overall, these patterns highlight a trade-off between \textit{query translation}, \textit{data quality}, and \textit{semantic alignment}. AMAZON queries fail due to diverse terminology and units, MAG queries due to complex content that is poorly captured by structured graph traversal, and PRIME queries due to navigating its rich schema.

\section{Related Work}
\label{sec:related_work}
Retrieval-augmented QA over heterogeneous corpora must handle not only textual relevance but also \emph{structure}: tables have schema and join constraints, documents have latent topical links, and many real collections expose explicit relations. Below, we summarize different techniques:

\paragraph{Table-centric retrieval and dataset discovery}
For table repositories, a major theme is retrieving \emph{related} tables using structural criteria such as unionability and joinability. Starmie \cite{fan2023starmie}, DeepJoin \cite{dong2023deepjoin}, and WarpGate \cite{cong2023warpgate} learn or encode column/table representations for efficient join or union search, typically accelerated via ANN indexing. Other work targets specialized relational augmentation signals, e.g., correlated dataset search via sketch-based indexing \cite{qcr2022icde}. For end-to-end table QA, systems improve practical retrieval quality with cascaded pipelines that combine sparse filtering, dense retrieval, and reranking \cite{singh2025craft}, while LLM-based dense retrieval techniques (e.g., HyDE) can improve zero-shot retrieval without labeled relevance data.

\paragraph{Graph-based indexing for retrieval and RAG}
Graph-based RAG methods organize information as graphs to retrieve connected context and aggregate evidence. Microsoft GraphRAG \cite{edge2024graphrag} builds an entity graph with community summaries for corpus-level questions; GRAG \cite{hu2024grag} retrieves textual subgraphs and injects topology into generation; and PathRAG \cite{chen2025pathrag} focuses on redundancy pruning and path-based prompting. GNN-Ret \cite{li2024graphneuralnetworkenhanced} constructs passage graphs using structural (same section/document) and keyword edges, then applies a GNN to re-rank results without semantic chunking or metadata pruning. ATLANTIC \cite{munikoti2023atlanticstructureawareretrievalaugmentedlanguage} builds document-level graphs with predefined structural relations and encodes them with a GNN, operating at the document rather than chunk level. In multi-document QA, knowledge-graph prompting assembles supporting context via passage graph traversal \cite{wang2023knowledgegraphpromptingmultidocument}. A broader survey \cite{han2025graphrag} notes that graph form and domain constraints strongly shape the design space.

\paragraph{LM+KG reasoning and agentic retrieval with tools}
When explicit KGs are available, models such as QA-GNN \cite{wang2024knowledge, sun2018opendomainquestionanswering, sun2019pullnetopendomainquestion, yasunaga2022qagnnreasoninglanguagemodels, Pan_2024} combine LM-based relevance estimation with graph neural reasoning. In parallel, agentic frameworks treat retrieval as a plan--execute process over external tools: ReAct interleaves reasoning and actions, Reflexion  improves agents via feedback and self-reflection, AvaTaR  optimizes tool use via contrastive reasoning.

Prior work often targets a single structure type (table repositories, explicit KGs, or graph indices for text corpora). Our framework is retrieval-centric and structure-aware: we build chunk graphs offline when relations are implicit, preserve native relations when graphs are explicit, and use a simple \emph{seed $\rightarrow$ expand $\rightarrow$ rerank} operator to select evidence under a fixed budget.

\section{Conclusion}
We present \methodname, a structure-aware graph retrieval framework that adapts retrieval operators to data topology by separating offline graph construction from online query-conditioned traversal. Metadata-driven similarity and pruning build graphs for implicit corpora, while native schema edges are preserved for explicit KGs, with operators selected at query time based on graph density and schema availability.

On OTT-QA, similarity-driven expansion boosts Recall@$k$ by 2.8-5.7 points over strong hybrid baselines, retrieving more coherent multi-hop evidence with compact context. On STaRK, agentic retrieval with selective expansion approaches fine-tuned performance without training, using symbolic Cypher queries to navigate dense graphs while limiting neighbor noise. Overall, structure-aware retrieval that leverages graph connectivity can match or outperform strong hybrid baselines without task-specific training.

Future work may include instruction tuning or fine-tuning the agent to generate more accurate and efficient Cypher queries, as well as extending the framework to multimodal graphs that incorporate images and other media through cross-modal links.

\section*{Limitations}
Our study focuses on two representative regimes and relies on LLM-based agents for STaRK, which introduces latency and computational cost compared to traditional retrieval. This tradeoff enables training-free symbolic Cypher generation that provides precise schema-aware traversal while avoiding neighbor noise in dense graphs. The framework depends on initial seed quality, as graph expansion amplifies rather than replaces base retrieval. Error analysis reveals constraint extraction and filtering (30 to 36 percent of errors) as the dominant failure mode, reflecting the difficulty of mapping informal queries to structured predicates without task-specific training. Despite these limitations, we excel at surfacing structurally connected multi-hop evidence that flat retrieval misses, achieving 2.8 to 5.7 point gains on OTT-QA and approaching fine-tuned baselines on STaRK without training, with particularly strong performance on complex queries where graph topology provides discriminative structure.

\section*{Ethical Statement}
We evaluate on publicly available benchmarks (OTT-QA and STaRK) released for research use under their respective licenses. Our framework operates on graph structures and does not collect any new user data or infer personal or demographic attributes. 

Importantly, our agentic approach accesses only graph topology and schema information rather than actual node content during query planning, which provides an additional layer of privacy protection and reduces exposure to sensitive information. We aim to benefit the research community by enabling more efficient structure-aware retrieval across heterogeneous information sources.

As with any retrieval system, our approach could be misused in real deployments to surface or combine information without authorization. However, our work is intended solely for academic research and is not designed for deployment in surveillance, decision-making, or other high-stakes settings. Any practical use should follow standard data-governance practices (access control, auditing, and privacy safeguards) and undergo appropriate oversight. To support reproducibility, we document dataset splits, prompts, and decoding settings. 

We used AI tools to assist with the writing process and improve the presentation.

\section*{Acknowledgement }
This research has been supported in part by the ONR Contract N00014-23-1-2364, and conducted as a collaborative effort between Arizona State University and the University of Pennsylvania. We gratefully acknowledge the Complex Data Analysis and Reasoning Lab at School of Augmented Intelligence, Arizona State University and Cognitive Computation Group, University of Pennsylvania for providing computational resources and institutional support.

\bibliographystyle{acl_natbib}
\bibliography{custom} 

\begin{thebibliography}{53}
\expandafter\ifx\csname natexlab\endcsname\relax\def\natexlab#1{#1}\fi

\bibitem[{201(2013)}]{2013sparql}
 2013.
\newblock \href {http://www.w3.org/TR/sparql11-query} {{SPARQL 1.1 Query Language}}.
\newblock Technical report, W3C.

\bibitem[{Arabzadeh et~al.(2021)Arabzadeh, Yan, and Clarke}]{arabzadeh2021predicting}
Negar Arabzadeh, Xinyi Yan, and Charles~LA Clarke. 2021.
\newblock Predicting efficiency/effectiveness trade-offs for dense vs. sparse retrieval strategy selection.
\newblock In \emph{Proceedings of the 30th ACM International Conference on Information \& Knowledge Management}, pages 2862--2866.

\bibitem[{Boer et~al.(2024)Boer, Koch, and Kramer}]{Boe24}
Derian Boer, Fabian Koch, and Stefan Kramer. 2024.
\newblock Harnessing the power of semi-structured knowledge and llms with triplet-based prefiltering for question answering.
\newblock \emph{arXiv preprint arXiv:2409.00861}.

\bibitem[{Boer et~al.(2025)Boer, Roth, and Kramer}]{boer2025focus}
Derian Boer, Stephen Roth, and Stefan Kramer. 2025.
\newblock Focus, merge, rank: Improved question answering based on semi-structured knowledge bases.
\newblock \emph{arXiv preprint arXiv:2505.09246}.

\bibitem[{Chen et~al.(2025{\natexlab{a}})Chen, Guo, Yang, Chen, Chen, Liu, Shi, and Yang}]{chen2025pathrag}
Boyu Chen, Zirui Guo, Zidan Yang, Yuluo Chen, Junze Chen, Zhenghao Liu, Chuan Shi, and Cheng Yang. 2025{\natexlab{a}}.
\newblock Pathrag: Pruning graph-based retrieval augmented generation with relational paths.
\newblock \emph{arXiv preprint arXiv:2502.14902}.

\bibitem[{Chen et~al.(2025{\natexlab{b}})Chen, Zhang, Cafarella, and Roth}]{chen-etal-2025-retrieve}
Peter~Baile Chen, Yi~Zhang, Mike Cafarella, and Dan Roth. 2025{\natexlab{b}}.
\newblock \href {https://doi.org/10.18653/v1/2025.acl-long.1463} {Can we retrieve everything all at once? {ARM}: An alignment-oriented {LLM}-based retrieval method}.
\newblock In \emph{Proceedings of the 63rd Annual Meeting of the Association for Computational Linguistics (Volume 1: Long Papers)}, pages 30298--30317, Vienna, Austria. Association for Computational Linguistics.

\bibitem[{Chen et~al.(2020)Chen, Chang, Schlinger, Wang, and Cohen}]{chen2020open}
Wenhu Chen, Ming-Wei Chang, Eva Schlinger, William Wang, and William~W Cohen. 2020.
\newblock Open question answering over tables and text.
\newblock \emph{arXiv preprint arXiv:2010.10439}.

\bibitem[{Chen et~al.(2022)Chen, Lakhotia, Oguz, Gupta, Lewis, Peshterliev, Mehdad, Gupta, and Yih}]{chen2022salient}
Xilun Chen, Kushal Lakhotia, Barlas Oguz, Anchit Gupta, Patrick Lewis, Stan Peshterliev, Yashar Mehdad, Sonal Gupta, and Wen-tau Yih. 2022.
\newblock Salient phrase aware dense retrieval: can a dense retriever imitate a sparse one?
\newblock In \emph{Findings of the Association for Computational Linguistics: EMNLP 2022}, pages 250--262.

\bibitem[{Chen et~al.(2024)Chen, Chen, He, Wen, and Sun}]{chen-etal-2024-analyze}
Xinran Chen, Xuanang Chen, Ben He, Tengfei Wen, and Le~Sun. 2024.
\newblock \href {https://doi.org/10.18653/v1/2024.findings-acl.708} {Analyze, generate and refine: Query expansion with {LLM}s for zero-shot open-domain {QA}}.
\newblock In \emph{Findings of the Association for Computational Linguistics: ACL 2024}, pages 11908--11922, Bangkok, Thailand. Association for Computational Linguistics.

\bibitem[{Cong et~al.(2022)Cong, Gale, Frantz, Jagadish, and Demiralp}]{cong2023warpgate}
Tianji Cong, James Gale, Jason Frantz, HV~Jagadish, and {\c{C}}a{\u{g}}atay Demiralp. 2022.
\newblock Warpgate: A semantic join discovery system for cloud data warehouses.
\newblock \emph{arXiv preprint arXiv:2212.14155}.

\bibitem[{Dong et~al.(2022)Dong, Xiao, Nozawa, Enomoto, and Oyamada}]{dong2023deepjoin}
Yuyang Dong, Chuan Xiao, Takuma Nozawa, Masafumi Enomoto, and Masafumi Oyamada. 2022.
\newblock Deepjoin: Joinable table discovery with pre-trained language models.
\newblock \emph{arXiv preprint arXiv:2212.07588}.

\bibitem[{Edge et~al.(2024{\natexlab{a}})Edge, Trinh, Cheng, Bradley, Chao, Mody, Truitt, Metropolitansky, Ness, and Larson}]{edge2024local}
Darren Edge, Ha~Trinh, Newman Cheng, Joshua Bradley, Alex Chao, Apurva Mody, Steven Truitt, Dasha Metropolitansky, Robert~Osazuwa Ness, and Jonathan Larson. 2024{\natexlab{a}}.
\newblock From local to global: A graph rag approach to query-focused summarization.
\newblock \emph{arXiv preprint arXiv:2404.16130}.

\bibitem[{Edge et~al.(2024{\natexlab{b}})Edge, Trinh, Cheng, Bradley, Chao, Mody, Truitt, Metropolitansky, Ness, and Larson}]{edge2024graphrag}
Darren Edge, Ha~Trinh, Newman Cheng, Joshua Bradley, Alex Chao, Apurva Mody, Steven Truitt, Dasha Metropolitansky, Robert~Osazuwa Ness, and Jonathan Larson. 2024{\natexlab{b}}.
\newblock From local to global: A graph rag approach to query-focused summarization.
\newblock \emph{arXiv preprint arXiv:2404.16130}.

\bibitem[{Fan et~al.(2022)Fan, Wang, Li, Zhang, and Miller}]{fan2023starmie}
Grace Fan, Jin Wang, Yuliang Li, Dan Zhang, and Ren{\'e}e Miller. 2022.
\newblock Semantics-aware dataset discovery from data lakes with contextualized column-based representation learning.
\newblock \emph{arXiv preprint arXiv:2210.01922}.

\bibitem[{Francis et~al.(2018)Francis, Green, Guagliardo, Libkin, Lindaaker, Marsault, Plantikow, Rydberg, Selmer, and Taylor}]{Fra18}
Nadime Francis, Alastair Green, Paolo Guagliardo, Leonid Libkin, Tobias Lindaaker, Victor Marsault, Stefan Plantikow, Mats Rydberg, Petra Selmer, and Andr{\'e}s Taylor. 2018.
\newblock Cypher: An evolving query language for property graphs.
\newblock In \emph{Proceedings of the 2018 international conference on management of data}, pages 1433--1445.

\bibitem[{Gao et~al.(2023)Gao, Ma, Lin, and Callan}]{gao2022hyde}
Luyu Gao, Xueguang Ma, Jimmy Lin, and Jamie Callan. 2023.
\newblock Precise zero-shot dense retrieval without relevance labels.
\newblock In \emph{Proceedings of the 61st Annual Meeting of the Association for Computational Linguistics (Volume 1: Long Papers)}, pages 1762--1777.

\bibitem[{Han et~al.(2024)Han, Wang, Shomer, Guo, Ding, Lei, Halappanavar, Rossi, Mukherjee, Tang et~al.}]{han2025graphrag}
Haoyu Han, Yu~Wang, Harry Shomer, Kai Guo, Jiayuan Ding, Yongjia Lei, Mahantesh Halappanavar, Ryan~A Rossi, Subhabrata Mukherjee, Xianfeng Tang, et~al. 2024.
\newblock Retrieval-augmented generation with graphs (graphrag).
\newblock \emph{arXiv preprint arXiv:2501.00309}.

\bibitem[{Huang et~al.(2022)Huang, Zhong, Liu, Gong, Jiang, and Duan}]{huang-etal-2022-mixed}
Junjie Huang, Wanjun Zhong, Qian Liu, Ming Gong, Daxin Jiang, and Nan Duan. 2022.
\newblock \href {https://doi.org/10.18653/v1/2022.findings-emnlp.303} {Mixed-modality representation learning and pre-training for joint table-and-text retrieval in {O}pen{QA}}.
\newblock In \emph{Findings of the Association for Computational Linguistics: EMNLP 2022}, pages 4117--4129, Abu Dhabi, United Arab Emirates. Association for Computational Linguistics.

\bibitem[{Jimenez~Gutierrez et~al.(2024)Jimenez~Gutierrez, Shu, Gu, Yasunaga, and Su}]{jimenez2024hipporag}
Bernal Jimenez~Gutierrez, Yiheng Shu, Yu~Gu, Michihiro Yasunaga, and Yu~Su. 2024.
\newblock Hipporag: Neurobiologically inspired long-term memory for large language models.
\newblock \emph{Advances in Neural Information Processing Systems}, 37:59532--59569.

\bibitem[{Karpukhin et~al.(2020)Karpukhin, Oguz, Min, Lewis, Wu, Edunov, Chen, and Yih}]{karpukhin2020dense}
Vladimir Karpukhin, Barlas Oguz, Sewon Min, Patrick~SH Lewis, Ledell Wu, Sergey Edunov, Danqi Chen, and Wen-tau Yih. 2020.
\newblock Dense passage retrieval for open-domain question answering.
\newblock In \emph{EMNLP (1)}, pages 6769--6781.

\bibitem[{Lei et~al.(2025)Lei, Han, Rossi, Dernoncourt, Lipka, Halappanavar, Tang, and Wang}]{Lei25}
Yongjia Lei, Haoyu Han, Ryan~A Rossi, Franck Dernoncourt, Nedim Lipka, Mahantesh~M Halappanavar, Jiliang Tang, and Yu~Wang. 2025.
\newblock Mixture of structural-and-textual retrieval over text-rich graph knowledge bases.
\newblock \emph{arXiv preprint arXiv:2502.20317}.

\bibitem[{Lewis et~al.(2020)Lewis, Perez, Piktus, Petroni, Karpukhin, Goyal, K{\"u}ttler, Lewis, Yih, Rockt{\"a}schel et~al.}]{lewis2020retrieval}
Patrick Lewis, Ethan Perez, Aleksandra Piktus, Fabio Petroni, Vladimir Karpukhin, Naman Goyal, Heinrich K{\"u}ttler, Mike Lewis, Wen-tau Yih, Tim Rockt{\"a}schel, et~al. 2020.
\newblock Retrieval-augmented generation for knowledge-intensive nlp tasks.
\newblock \emph{Advances in neural information processing systems}, 33:9459--9474.

\bibitem[{Li et~al.(2024{\natexlab{a}})Li, Zhang, and Liu}]{Li24}
Yihao Li, Ru~Zhang, and Jianyi Liu. 2024{\natexlab{a}}.
\newblock An enhanced prompt-based llm reasoning scheme via knowledge graph-integrated collaboration.
\newblock In \emph{International Conference on Artificial Neural Networks}, pages 251--265.

\bibitem[{Li et~al.(2024{\natexlab{b}})Li, Guo, Shao, Song, Bian, Zhang, and Wang}]{li2024graphneuralnetworkenhanced}
Zijian Li, Qingyan Guo, Jiawei Shao, Lei Song, Jiang Bian, Jun Zhang, and Rui Wang. 2024{\natexlab{b}}.
\newblock \href {http://arxiv.org/abs/2406.06572} {Graph neural network enhanced retrieval for question answering of llms}.

\bibitem[{Ma et~al.(2022)Ma, Cheng, Liu, Nyberg, and Gao}]{ma-etal-2022-open-domain}
Kaixin Ma, Hao Cheng, Xiaodong Liu, Eric Nyberg, and Jianfeng Gao. 2022.
\newblock \href {https://doi.org/10.18653/v1/2022.findings-emnlp.392} {Open-domain question answering via chain of reasoning over heterogeneous knowledge}.
\newblock In \emph{Findings of the Association for Computational Linguistics: EMNLP 2022}, pages 5360--5374, Abu Dhabi, United Arab Emirates. Association for Computational Linguistics.

\bibitem[{Ma et~al.(2023)Ma, Cheng, Zhang, Liu, Nyberg, and Gao}]{ma-etal-2023-chain}
Kaixin Ma, Hao Cheng, Yu~Zhang, Xiaodong Liu, Eric Nyberg, and Jianfeng Gao. 2023.
\newblock \href {https://doi.org/10.18653/v1/2023.acl-long.89} {Chain-of-skills: A configurable model for open-domain question answering}.
\newblock In \emph{Proceedings of the 61st Annual Meeting of the Association for Computational Linguistics (Volume 1: Long Papers)}, pages 1599--1618, Toronto, Canada. Association for Computational Linguistics.

\bibitem[{Malkov and Yashunin(2018)}]{malkov2018efficient}
Yu~A Malkov and Dmitry~A Yashunin. 2018.
\newblock Efficient and robust approximate nearest neighbor search using hierarchical navigable small world graphs.
\newblock \emph{IEEE transactions on pattern analysis and machine intelligence}, 42(4):824--836.

\bibitem[{Mandikal and Mooney(2024)}]{mandikal2024sparse}
Priyanka Mandikal and Raymond Mooney. 2024.
\newblock Sparse meets dense: A hybrid approach to enhance scientific document retrieval.
\newblock \emph{arXiv preprint arXiv:2401.04055}.

\bibitem[{Mavromatis and Karypis(2024)}]{mavromatis2024gnn}
Costas Mavromatis and George Karypis. 2024.
\newblock Gnn-rag: Graph neural retrieval for large language model reasoning.
\newblock \emph{arXiv preprint arXiv:2405.20139}.

\bibitem[{Munikoti et~al.(2023)Munikoti, Acharya, Wagle, and Horawalavithana}]{munikoti2023atlanticstructureawareretrievalaugmentedlanguage}
Sai Munikoti, Anurag Acharya, Sridevi Wagle, and Sameera Horawalavithana. 2023.
\newblock \href {http://arxiv.org/abs/2311.12289} {Atlantic: Structure-aware retrieval-augmented language model for interdisciplinary science}.

\bibitem[{{Neo4j, Inc.}(2026)}]{neo4j2026}
{Neo4j, Inc.} 2026.
\newblock \href {https://neo4j.com/} {\emph{Neo4j Graph Database}}.

\bibitem[{OpenAI et~al.(2024)OpenAI, :, Hurst, Lerer, Goucher, Perelman, Ramesh, Clark, Ostrow, Welihinda, and et~al.}]{gpt4ocard}
OpenAI, :, Aaron Hurst, Adam Lerer, Adam~P. Goucher, Adam Perelman, Aditya Ramesh, Aidan Clark, AJ~Ostrow, Akila Welihinda, and et~al. 2024.
\newblock \href {http://arxiv.org/abs/2410.21276} {Gpt-4o system card}.

\bibitem[{Pan et~al.(2024)Pan, Luo, Wang, Chen, Wang, and Wu}]{Pan_2024}
Shirui Pan, Linhao Luo, Yufei Wang, Chen Chen, Jiapu Wang, and Xindong Wu. 2024.
\newblock Unifying large language models and knowledge graphs: A roadmap.
\newblock \emph{IEEE Transactions on Knowledge and Data Engineering}, 36(7):3580--3599.

\bibitem[{Park et~al.(2023)Park, min Lee, Seo, Kim, Kang, and Na}]{Park2023RINKRE}
Eunhwan Park, Sung min Lee, Dearyong Seo, Seonhoon Kim, Inho Kang, and Seung-Hoon Na. 2023.
\newblock \href {https://api.semanticscholar.org/CorpusID:259549340} {Rink: Reader-inherited evidence reranker for table-and-text open domain question answering}.
\newblock In \emph{AAAI Conference on Artificial Intelligence}.

\bibitem[{Peng et~al.(2024)Peng, Zhu, Liu, Bo, Shi, Hong, Zhang, and Tang}]{hu2024grag}
Boci Peng, Yun Zhu, Yongchao Liu, Xiaohe Bo, Haizhou Shi, Chuntao Hong, Yan Zhang, and Siliang Tang. 2024.
\newblock Graph retrieval-augmented generation: A survey.
\newblock \emph{ACM Transactions on Information Systems}.

\bibitem[{Reimers and Gurevych(2019)}]{reimers2019sentence}
Nils Reimers and Iryna Gurevych. 2019.
\newblock Sentence-bert: Sentence embeddings using siamese bert-networks.
\newblock \emph{arXiv preprint arXiv:1908.10084}.

\bibitem[{Robertson et~al.(2009)Robertson, Zaragoza et~al.}]{robertson2009probabilistic}
Stephen Robertson, Hugo Zaragoza, et~al. 2009.
\newblock The probabilistic relevance framework: Bm25 and beyond.
\newblock \emph{Foundations and Trends{\textregistered} in Information Retrieval}, 3(4):333--389.

\bibitem[{Santos et~al.(2022)Santos, Bessa, Musco, and Freire}]{qcr2022icde}
Aécio Santos, Aline Bessa, Christopher Musco, and Juliana Freire. 2022.
\newblock \href {https://doi.org/10.1109/ICDE53745.2022.00264} {A sketch-based index for correlated dataset search}.
\newblock In \emph{2022 IEEE 38th International Conference on Data Engineering (ICDE)}, pages 2928--2941.

\bibitem[{Shen et~al.(2024)Shen, Long, Geng, Tao, Lei, Zhou, Blumenstein, and Jiang}]{shen-etal-2024-retrieval}
Tao Shen, Guodong Long, Xiubo Geng, Chongyang Tao, Yibin Lei, Tianyi Zhou, Michael Blumenstein, and Daxin Jiang. 2024.
\newblock \href {https://doi.org/10.18653/v1/2024.findings-acl.943} {Retrieval-augmented retrieval: Large language models are strong zero-shot retriever}.
\newblock In \emph{Findings of the Association for Computational Linguistics: ACL 2024}, pages 15933--15946, Bangkok, Thailand. Association for Computational Linguistics.

\bibitem[{Shinn et~al.(2023)Shinn, Cassano, Gopinath, Narasimhan, and Yao}]{shinn2023reflexion}
Noah Shinn, Federico Cassano, Ashwin Gopinath, Karthik Narasimhan, and Shunyu Yao. 2023.
\newblock Reflexion: Language agents with verbal reinforcement learning.
\newblock \emph{Advances in Neural Information Processing Systems}, 36:8634--8652.

\bibitem[{Singh et~al.(2025)Singh, Bhandari, Gao, Dan, and Gupta}]{singh2025craft}
Adarsh Singh, Kushal~Raj Bhandari, Jianxi Gao, Soham Dan, and Vivek Gupta. 2025.
\newblock Craft: Training-free cascaded retrieval for tabular qa.
\newblock \emph{arXiv preprint arXiv:2505.14984}.

\bibitem[{Sun et~al.(2019)Sun, Bedrax-Weiss, and Cohen}]{sun2019pullnetopendomainquestion}
Haitian Sun, Tania Bedrax-Weiss, and William Cohen. 2019.
\newblock In \emph{Proceedings of the 2019 conference on empirical methods in natural language processing and the 9th international joint conference on natural language processing (EMNLP-IJCNLP)}, pages 2380--2390.

\bibitem[{Sun et~al.(2018)Sun, Dhingra, Zaheer, Mazaitis, Salakhutdinov, and Cohen}]{sun2018opendomainquestionanswering}
Haitian Sun, Bhuwan Dhingra, Manzil Zaheer, Kathryn Mazaitis, Ruslan Salakhutdinov, and William Cohen. 2018.
\newblock Open domain question answering using early fusion of knowledge bases and text.
\newblock In \emph{Proceedings of the 2018 conference on empirical methods in natural language processing}, pages 4231--4242.

\bibitem[{Wang et~al.(2020)Wang, Wei, Dong, Bao, Yang, and Zhou}]{wang2020minilm}
Wenhui Wang, Furu Wei, Li~Dong, Hangbo Bao, Nan Yang, and Ming Zhou. 2020.
\newblock Minilm: Deep self-attention distillation for task-agnostic compression of pre-trained transformers.
\newblock \emph{Advances in neural information processing systems}, 33:5776--5788.

\bibitem[{Wang et~al.(2023)Wang, Lipka, Rossi, Siu, Zhang, and Derr}]{wang2023knowledgegraphpromptingmultidocument}
Yu~Wang, Nedim Lipka, Ryan~A. Rossi, Alexa Siu, Ruiyi Zhang, and Tyler Derr. 2023.
\newblock \href {http://arxiv.org/abs/2308.11730} {Knowledge graph prompting for multi-document question answering}.

\bibitem[{Wang et~al.(2024)Wang, Lipka, Rossi, Siu, Zhang, and Derr}]{wang2024knowledge}
Yu~Wang, Nedim Lipka, Ryan~A Rossi, Alexa Siu, Ruiyi Zhang, and Tyler Derr. 2024.
\newblock Knowledge graph prompting for multi-document question answering.
\newblock In \emph{Proceedings of the AAAI conference on artificial intelligence}, volume~38, pages 19206--19214.

\bibitem[{Wu et~al.(2024{\natexlab{a}})Wu, Zhao, Huang, Huang, Yasunaga, Cao, Ioannidis, Subbian, Leskovec, and Zou}]{wu2024avatar}
Shirley Wu, Shiyu Zhao, Qian Huang, Kexin Huang, Michihiro Yasunaga, Kaidi Cao, Vassilis Ioannidis, Karthik Subbian, Jure Leskovec, and James~Y Zou. 2024{\natexlab{a}}.
\newblock Avatar: Optimizing llm agents for tool usage via contrastive reasoning.
\newblock \emph{Advances in Neural Information Processing Systems}, 37:25981--26010.

\bibitem[{Wu et~al.(2024{\natexlab{b}})Wu, Zhao, Yasunaga, Huang, Cao, Huang, Ioannidis, Subbian, Zou, and Leskovec}]{wu24}
Shirley Wu, Shiyu Zhao, Michihiro Yasunaga, Kexin Huang, Kaidi Cao, Qian Huang, Vassilis~N. Ioannidis, Karthik Subbian, James Zou, and Jure Leskovec. 2024{\natexlab{b}}.
\newblock Stark: Benchmarking llm retrieval on textual and relational knowledge bases.
\newblock In \emph{NeurIPS Datasets and Benchmarks Track}.

\bibitem[{Xia et~al.(2024)Xia, Wu, Kim, Yu, Rossi, Wang, and McAuley}]{Xia24}
Yu~Xia, Junda Wu, Sungchul Kim, Tong Yu, Ryan~A Rossi, Haoliang Wang, and Julian McAuley. 2024.
\newblock Knowledge-aware query expansion with large language models for textual and relational retrieval.
\newblock \emph{arXiv preprint arXiv:2410.13765}.

\bibitem[{Yao et~al.(2022)Yao, Zhao, Yu, Du, Shafran, Narasimhan, and Cao}]{yao2023reactsynergizingreasoningacting}
Shunyu Yao, Jeffrey Zhao, Dian Yu, Nan Du, Izhak Shafran, Karthik~R Narasimhan, and Yuan Cao. 2022.
\newblock React: Synergizing reasoning and acting in language models.
\newblock In \emph{The eleventh international conference on learning representations}.

\bibitem[{Yasunaga et~al.(2021)Yasunaga, Ren, Bosselut, Liang, and Leskovec}]{yasunaga2022qagnnreasoninglanguagemodels}
Michihiro Yasunaga, Hongyu Ren, Antoine Bosselut, Percy Liang, and Jure Leskovec. 2021.
\newblock Qa-gnn: Reasoning with language models and knowledge graphs for question answering.
\newblock \emph{arXiv preprint arXiv:2104.06378}.

\bibitem[{Zhang et~al.(2024)Zhang, Liu, Zhu, Zeng, Sheng, Yang, Dai, and Wang}]{zhang2024efficient}
Haoyu Zhang, Jun Liu, Zhenhua Zhu, Shulin Zeng, Maojia Sheng, Tao Yang, Guohao Dai, and Yu~Wang. 2024.
\newblock Efficient and effective retrieval of dense-sparse hybrid vectors using graph-based approximate nearest neighbor search.
\newblock \emph{arXiv preprint arXiv:2410.20381}.

\bibitem[{Zhong et~al.(2022)Zhong, Huang, Liu, Zhou, Wang, Yin, and Duan}]{zhong2022reasoning}
Wanjun Zhong, Junjie Huang, Qian Liu, Ming Zhou, Jiahai Wang, Jian Yin, and Nan Duan. 2022.
\newblock Reasoning over hybrid chain for table-and-text open domain qa.
\newblock \emph{arXiv preprint arXiv:2201.05880}.

\end{thebibliography}

\appendix

\section{Dataset Statistics}\label{sec:dataset_stats}

\subsection{OTT-QA}
OTT-QA consists of a heterogeneous graph connecting document and table chunks (Table~\ref{tab:ottqa_dataset_stats}). While table chunks are relatively few, they are densely linked to documents, enabling cross-modal retrieval between structured and unstructured sources. The large number of document-document edges reflects strong topical overlap across documents, which is particularly useful for graph-based context propagation.

\begin{table}[h]
\small
\centering
\setlength{\tabcolsep}{6pt}
\begin{tabular}{lc}
\toprule
\textbf{Component} & \textbf{Count} \\
\midrule
Document chunks & 21,912 \\
Table chunks & 974 \\
\midrule
\textbf{Total Nodes} & \textbf{22,886} \\
\midrule
Table--Table edges & 66,003 \\
Document--Table edges & 162,374 \\
Document--Document edges & 523,459 \\
\midrule
\textbf{Total Edges} & \textbf{751,836} \\
\bottomrule
\end{tabular}
\caption{OTT-QA dataset statistics showing node types, edge types, and their counts.}
\label{tab:ottqa_dataset_stats}
\end{table}
\subsection{STaRK}

\subsubsection{AMAZON}
The AMAZON dataset originally contains four node types: \texttt{review}, \texttt{brand}, \texttt{color}, and \texttt{category}. The latter three (\texttt{brand}, \texttt{color}, \texttt{category}) can alternatively be represented as metadata attributes rather than separate nodes. We observe that all product nodes contain a \texttt{reviews} key that stores tabular review data. 

For the dense+sparse (hybrid) retrieval approach, we treat this reviews data as a table for retrieval and expansion (Table~\ref{tab:amazon_nodes_stats}). Since Neo4j is a graph database and does not natively support tabular structures like SQL databases, we convert each review into a separate node connected to its corresponding product node (Table~\ref{tab:amazon_nodes_stats_neo4j}).. This graph-native representation enables efficient querying in cases where we need to aggregate data (e.g., star ratings) across multiple reviews for the same product.

\begin{table}[H]
\small
\centering
\setlength{\tabcolsep}{6pt}
\begin{tabular}{lc}
\toprule
\textbf{Node Type} & \textbf{Count} \\
\midrule
Review & 957,192 \\
Product & 957,192 \\
\midrule
\textbf{Total} & \textbf{1,914,384} \\
\bottomrule
\end{tabular}
\caption{AMAZON node type distribution from STaRK dataset for Hybrid retrieval.}
\label{tab:amazon_nodes_stats}
\end{table}

\begin{table}[H]
\small
\centering
\setlength{\tabcolsep}{6pt}
\begin{tabular}{lc}
\toprule
\textbf{Node Type} & \textbf{Count} \\
\midrule
Review & 12,963,012 \\
Product & 957,192 \\
\midrule
\textbf{Total} & \textbf{13,920,204} \\
\bottomrule
\end{tabular}
\caption{AMAZON node type distribution from STaRK dataset for Neo4j queries.}
\label{tab:amazon_nodes_stats_neo4j}
\end{table}

\begin{table}[H]
\small
\centering
\setlength{\tabcolsep}{6pt}
\begin{tabular}{lc}
\toprule
\textbf{Edge Type} & \textbf{Count} \\
\midrule
HAS\_REVIEW & 12,963,012 \\
ALSO\_BUY & 1,663,562 \\
ALSO\_VIEW & 1,362,680 \\
IS\_SIMILAR & 24,886,992 \\
\midrule
\textbf{Total} & \textbf{40,876,246} \\
\bottomrule
\end{tabular}
\caption{AMAZON edge type distribution from STaRK dataset.}
\label{tab:amazon_edges_stats}
\end{table}

\subsubsection{MAG}
MAG represents a large scholarly graph with diverse entity types but comparatively limited textual richness per node (Tables~\ref{tab:mag_nodes_stats} and~\ref{tab:mag_edges_stats}). While citation and authorship edges provide strong structural signals, the effectiveness of semantic graph expansion is constrained by shorter textual descriptions.

\begin{table}[H]
\small
\centering
\setlength{\tabcolsep}{6pt}
\begin{tabular}{lc}
\toprule
\textbf{Node Type} & \textbf{Count} \\
\midrule
Author & 1,104,554 \\
Paper & 700,244 \\
Field & 59,469 \\
Institution & 8,701 \\
\midrule
\textbf{Total} & \textbf{1,872,968} \\
\bottomrule
\end{tabular}
\caption{MAG node type distribution from STaRK dataset.}
\label{tab:mag_nodes_stats}
\end{table}

\begin{table}[H]
\small
\centering
\setlength{\tabcolsep}{6pt}
\begin{tabular}{lc}
\toprule
\textbf{Edge Type} & \textbf{Count} \\
\midrule
CITES & 9,719,488 \\
HAS\_FIELD & 7,243,269 \\
AUTHORED & 6,768,883 \\
ASSOCIATED\_WITH & 167,482 \\
IS\_SIMILAR & 12,954,514 \\
\midrule
\textbf{Total} & \textbf{36,853,636} \\
\bottomrule
\end{tabular}
\caption{MAG edge type distribution from STaRK dataset.}
\label{tab:mag_edges_stats}
\end{table}

\subsubsection{PRIME}
PRIME is a highly structured biomedical graph with many fine-grained entity and relation types (Tables~\ref{tab:prime_nodes_stats} and~\ref{tab:prime_edges_stats}). Although the graph is rich in relational semantics, node-level textual content is often sparse, making effective graph expansion more sensitive to the quality of initial retrieval.

\begin{table}[H]
\small
\centering
\setlength{\tabcolsep}{5pt}
\begin{tabular}{lc}
\toprule
\textbf{Node Type} & \textbf{Count} \\
\midrule
BiologicalProcess & 28,642 \\
Gene & 27,671 \\
Disease & 17,080 \\
Phenotype & 15,311 \\
Anatomy & 14,035 \\
MolecularFunction & 11,169 \\
Drug & 7,957 \\
CellularComponent & 4,176 \\
Pathway & 2,516 \\
Exposure & 818 \\
\midrule
\textbf{Total} & \textbf{129,375} \\
\bottomrule
\end{tabular}
\caption{PRIME node type distribution from STaRK dataset.}
\label{tab:prime_nodes_stats}
\end{table}

\begin{table}[h]
\small
\centering
\setlength{\tabcolsep}{2pt}
\begin{tabular}{lc}
\toprule
\textbf{Edge Type} & \textbf{Count} \\
\midrule
EXPRESSED\_IN & 3,036,406 \\
SYNERGISTIC\_WITH & 2,672,628 \\
AFFILIATED\_WITH & 1,029,162 \\
INTERACTS\_WITH & 686,550 \\
INTERACTS\_WITH\_PROTEIN & 642,150 \\
PHENOTYPE\_PRESENT & 300,634 \\
PARENT\_OF & 281,744 \\
HAS\_SIDE\_EFFECT & 129,568 \\
CONTRAINDICATED\_FOR & 61,350 \\
NOT\_EXPRESSED\_IN & 39,774 \\
TARGETS & 32,760 \\
INDICATED\_FOR & 18,776 \\
INHIBITS\_ENZYME & 10,634 \\
TRANSPORTED\_BY & 6,184 \\
OFF\_LABEL\_USE & 5,136 \\
LINKED\_TO & 4,608 \\
PHENOTYPE\_ABSENT & 2,386 \\
ACTS\_AS\_CARRIER & 1,728 \\
IS\_SIMILAR & 963,104 \\
\midrule
\textbf{Total} & \textbf{9,825,282} \\
\bottomrule
\end{tabular}
\caption{PRIME edge type distribution from STaRK dataset.}
\label{tab:prime_edges_stats}
\end{table}

\section{OTTQA graph building statistics}\label{sec:ott_graph_building_stats}

\subsection{Metadata Extraction Cost Analysis}

\begin{table}[H]
\small
\centering
\setlength{\tabcolsep}{6pt}
\begin{tabular}{lrrr}
\toprule
\textbf{Modality} & \textbf{Input} & \textbf{Output} & \textbf{Total} \\
\midrule
Table & 1,625,516 & 3,989,504 & 5,615,020 \\
Text & 53,501,851 & 21,960,752 & 75,462,603 \\
\midrule
\textbf{Grand Total} & \textbf{55,127,367} & \textbf{25,950,256} & \textbf{81,077,623} \\
\bottomrule
\end{tabular}
\caption{Token usage by modality for OTT-QA metadata extraction.}
\label{tab:ottqa_tokens}
\end{table}

As shown in Table~\ref{tab:ottqa_tokens}, text metadata extraction dominates total token usage, accounting for the majority of the 81.1M tokens processed. Using gpt-4o-mini pricing, table metadata extraction costs approximately \textit{\$2.64}, while text metadata extraction costs \textit{\$21.20}, for a combined total of \textbf{\$23.84}, \textbf{\$11.92} if we apply batching. Despite the large aggregate volume, the per-item processing cost remains extremely low, highlighting the cost efficiency of large-scale metadata generation. Open source models can also be used.

\subsection{Table-Table Similarity Analysis}

We analyze similarity statistics between table chunks in OTT-QA to evaluate whether LLMs are necessary for graph construction.

\begin{figure}[H]
\centering
\includegraphics[width=0.48\linewidth]{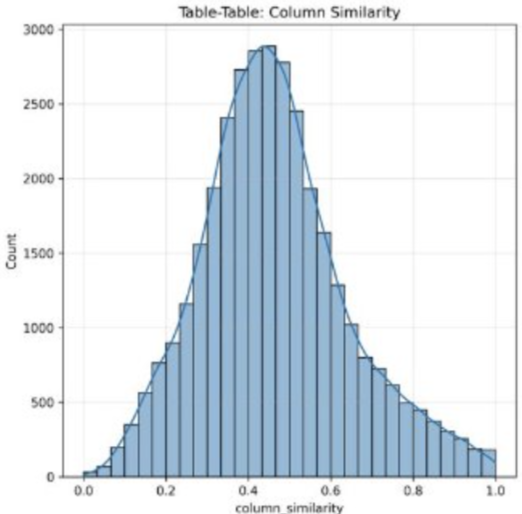}
\includegraphics[width=0.48\linewidth]{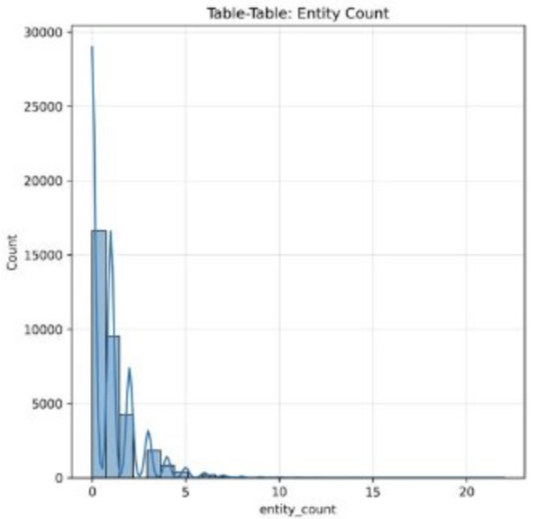}
\caption{\small Left: column similarity distribution. Right: number of entities per table chunk (left-skewed).}
\label{fig:tab_tab_col_sim}
\label{fig:tab_tab_entity}
\end{figure}

\paragraph{Column similarity and entity distribution}
Figure~\ref{fig:tab_tab_col_sim}  show the distribution of column similarity and number of entities per chunk. Column similarity is roughly bell-shaped (normal distribution), while entity counts are strongly left-skewed - most chunks contain only one entity. These figures highlight that high semantic similarity exists even for single-entity chunks.

\begin{figure}[H]
\centering
\includegraphics[width=0.48\linewidth]{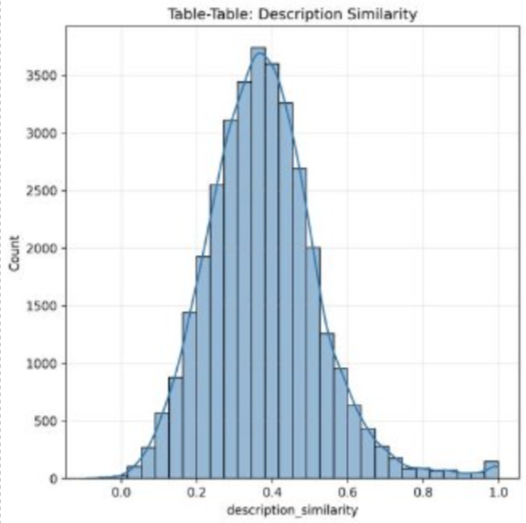}
\includegraphics[width=0.48\linewidth]{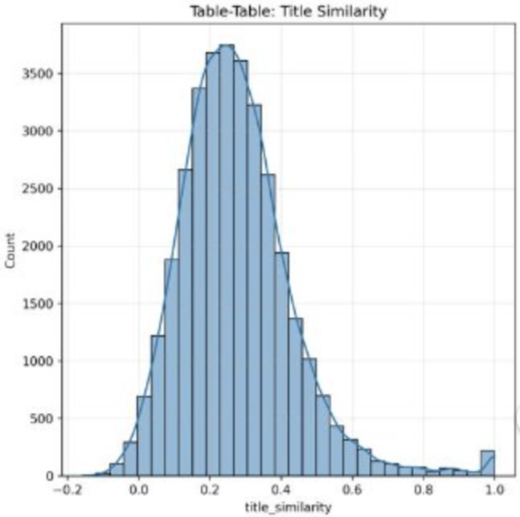}
\caption{\small Left: description similarity. Right: title similarity. Both distributions are similar, with a slight rise at high similarity.}
\label{fig:tab_tab_desc_sim}
\label{fig:tab_tab_title_sim}
\end{figure}

\paragraph{Description and title similarity}
Figure~\ref{fig:tab_tab_desc_sim} show the distributions of description and title similarity. Both are bell-shaped with a slight high-end tail, indicating that some tables are highly similar in title or description.

\begin{figure}[H]
\centering
\includegraphics[width=0.32\linewidth]{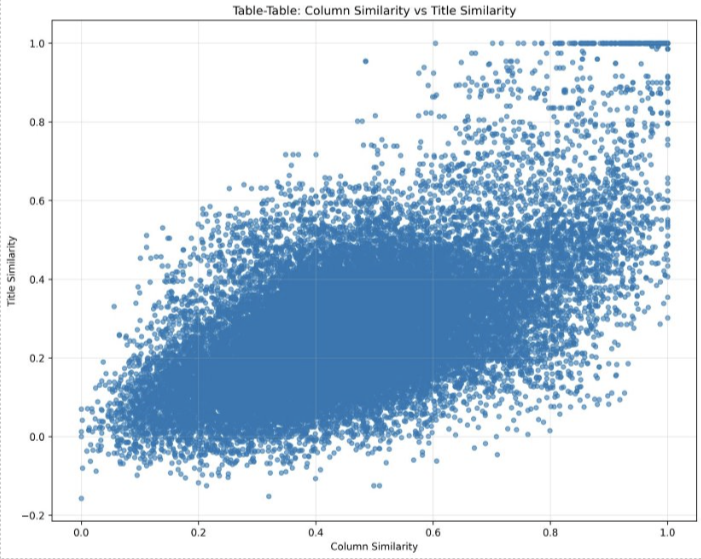}
\includegraphics[width=0.32\linewidth]{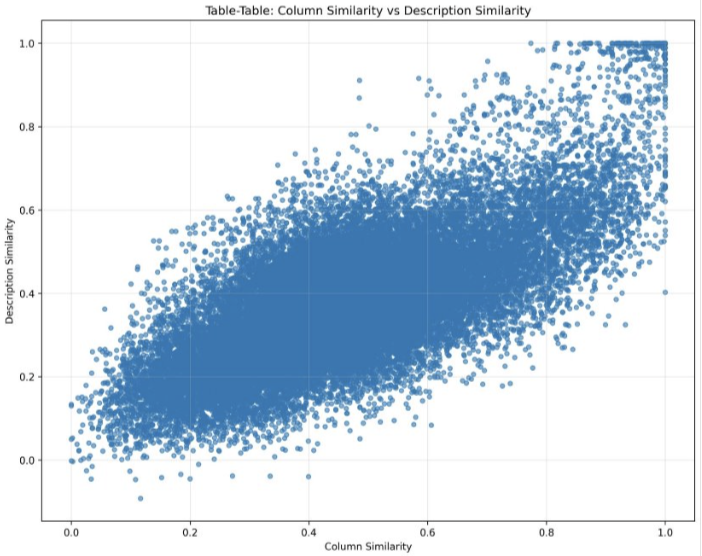}
\includegraphics[width=0.32\linewidth]{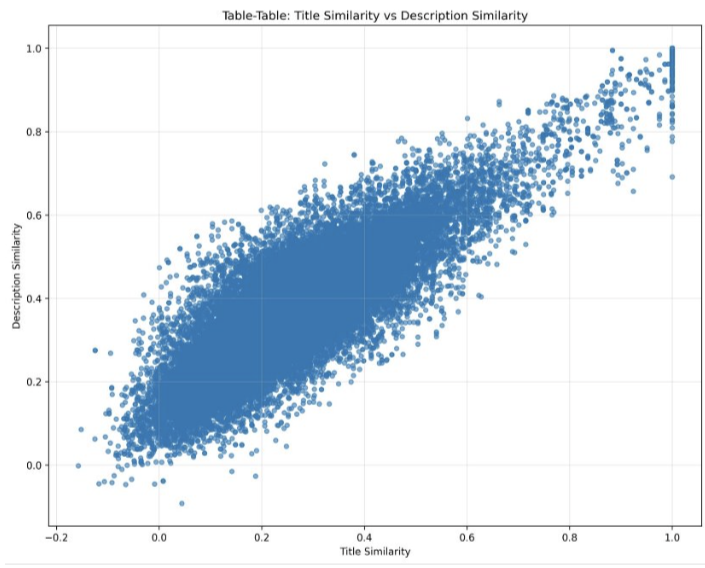}
\caption{\small Scatter plots of column-title (left), column-description (center), and title-description (right) similarities. Diagonal trends indicate strong alignment.}
\label{fig:tab_tab_col_title}
\label{fig:tab_tab_col_desc}
\label{fig:tab_tab_title_desc}
\end{figure}

\paragraph{Column-title, column-description, title-description scatter plots}
Figure~\ref{fig:tab_tab_col_title} show scatter plots of column-title, column-description, and title-description similarities, respectively. Strong diagonal trends indicate that similar columns tend to co-occur with similar titles and descriptions.

\paragraph{Correlation matrix}
Table~\ref{tab:tab_tab_matrix} shows the correlation matrix of all table-level similarity metrics. Column, title, and description similarities are strongly correlated, with title and description similarity correlation 0.846. 

\begin{table}[H]
    \centering
    \small
    \begin{tabular}{lcccc}
        \toprule
        & \textbf{(1)} & \textbf{(2)} & \textbf{(3)} & \textbf{(4)} \\
        \midrule
        \textbf{(1) Col Sim}   & 1.000 & 0.165 & 0.577 & 0.696 \\
        \textbf{(2) Ent Cnt}   & 0.165 & 1.000 & 0.179 & 0.189 \\
        \textbf{(3) Titl Sim}  & 0.577 & 0.179 & 1.000 & 0.846 \\
        \textbf{(4) Desc Sim}  & 0.696 & 0.189 & 0.846 & 1.000 \\
        \bottomrule
    \end{tabular}
    \caption{Table-Table Correlation Matrix. Abbreviations: (1) column\_similarity, (2) entity\_count, (3) title\_similarity, and (4) description\_similarity.}
    \label{tab:tab_tab_matrix}
\end{table}

\subsection{Document-Document Similarity Analysis}

We analyze similarity statistics between document chunks in OTT-QA to assess semantic content and the need for LLM-based processing.

\begin{figure}[H]
\centering
\includegraphics[width=0.9\linewidth]{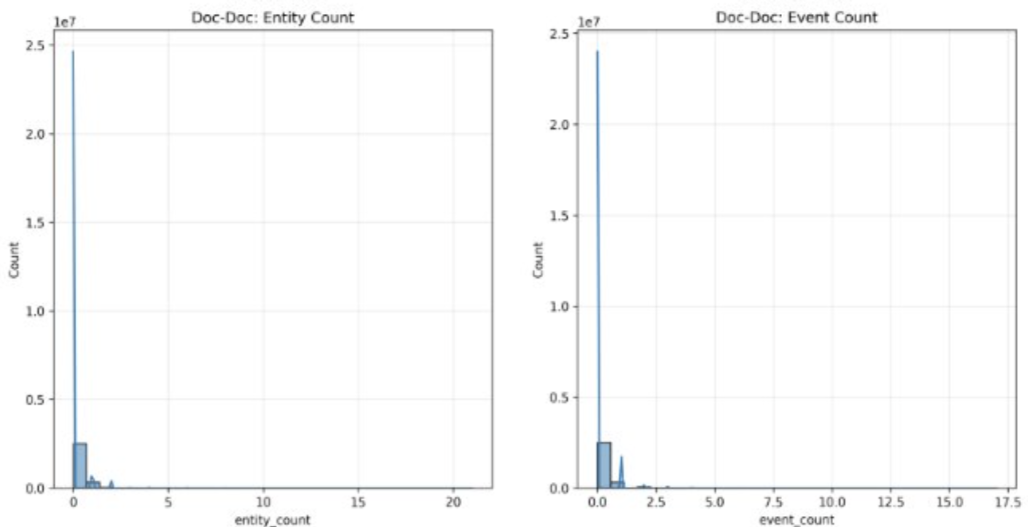}
\caption{\small Distribution of entities and events per document chunk. Most chunks contain only one entity, reducing the need for LLM-based extraction.}
\label{fig:doc_doc_entity_event}
\end{figure}

\paragraph{Entity and event counts per chunk}
Figure~\ref{fig:doc_doc_entity_event} shows the number of entities and events per document chunk. Both distributions are strongly left-skewed, with most chunks containing only a single entity or event. This indicates that document chunks are small and focused, and additional LLM-based extraction or summarization would provide minimal benefit.

\begin{figure}[H]
\centering
\includegraphics[width=0.9\linewidth]{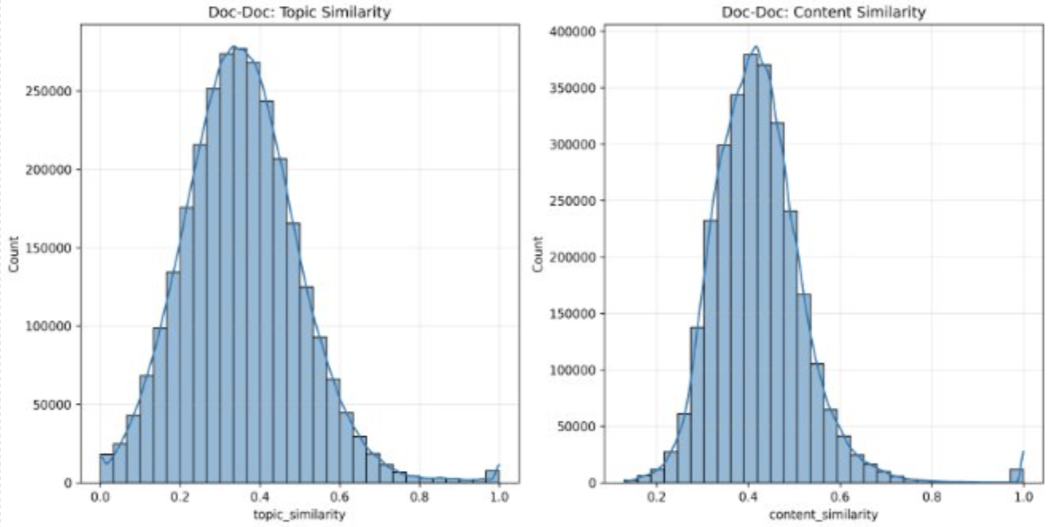}
\caption{\small Histograms of topic similarity (left) and content similarity (right) between document chunks. Bell-shaped distributions with a high-similarity spike are observed.}
\label{fig:doc_doc_topic_content}
\end{figure}

\paragraph{Topic and content similarity distributions}
Figure~\ref{fig:doc_doc_topic_content} shows histograms of topic similarity and content similarity between document chunks. Both distributions are roughly bell-shaped with a small spike near 1, indicating that while most document pairs are moderately similar, a few are nearly identical.

\begin{figure}[H]
\centering
\includegraphics[width=0.9\linewidth]{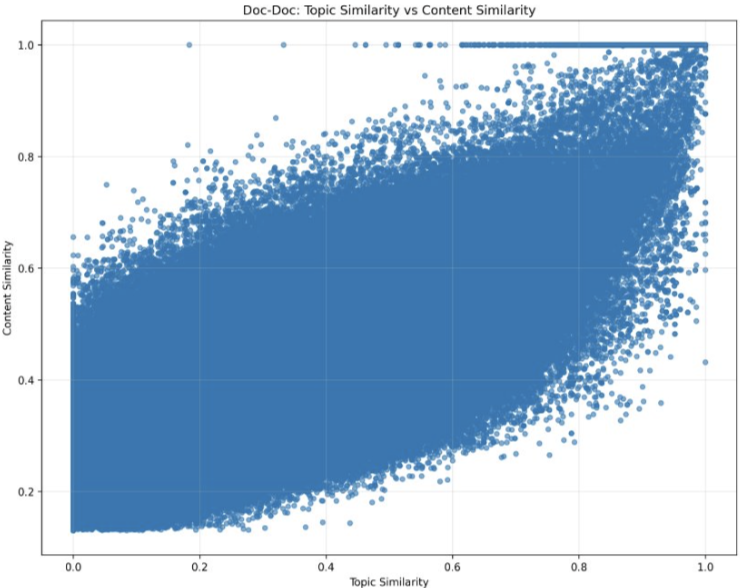}
\caption{\small Scatter plot of content similarity versus topic similarity. Diagonal concentration indicates strong alignment between the two metrics.}
\label{fig:doc_doc_content_topic}
\end{figure}

\paragraph{Topic-content scatter plot}
Figure~\ref{fig:doc_doc_content_topic} shows a scatter plot of content similarity versus topic similarity for each document pair. Most points lie along the diagonal, confirming strong alignment between content and topic similarities.

\paragraph{Correlation matrix}
Table~\ref{tab:doc_tab_matrix} shows the correlation matrix of document-table similarity metrics. Column similarity, title similarity, and summary similarity are moderately correlated, confirming that cross-modal semantic alignment can be captured using simple embeddings.

\begin{table}[H]
    \centering
    \footnotesize
    \setlength{\tabcolsep}{4pt}
    \begin{tabular}{lccccc}
        \toprule
        & \textbf{(1)} & \textbf{(2)} & \textbf{(3)} & \textbf{(4)} & \textbf{(5)} \\
        \midrule
        \textbf{(1) Top Sim}  & 1.000 & 0.650 & 0.250 & 0.253 & 0.153 \\
        \textbf{(2) Cont Sim} & 0.650 & 1.000 & 0.357 & 0.339 & 0.251 \\
        \textbf{(3) Ent Rel}  & 0.250 & 0.357 & 1.000 & 0.530 & 0.230 \\
        \textbf{(4) Ent Cnt}  & 0.253 & 0.339 & 0.530 & 1.000 & 0.235 \\
        \textbf{(5) Evt Cnt}  & 0.153 & 0.251 & 0.230 & 0.235 & 1.000 \\
        \bottomrule
    \end{tabular}
    \caption{Doc-Doc Correlation Matrix. Abbreviations: (1) topic\_similarity, (2) content\_similarity, (3) entity\_relationship\_overlap, (4) entity\_count, and (5) event\_count.}
    \label{tab:doc_doc_matrix}
\end{table}

\subsection{Document-Table Similarity Analysis}

We analyze the semantic alignment between document and table chunks in OTT-QA to evaluate whether LLM-based processing is necessary.

\begin{figure}[H]
\centering
\includegraphics[width=0.48\linewidth]{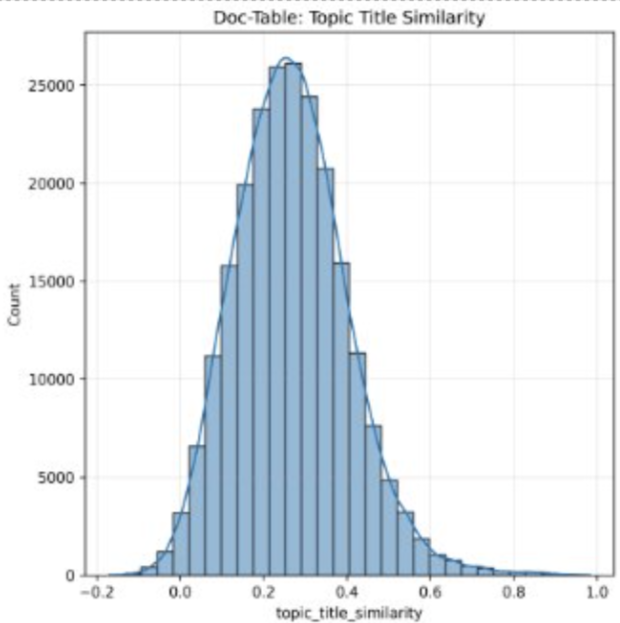}
\includegraphics[width=0.48\linewidth]{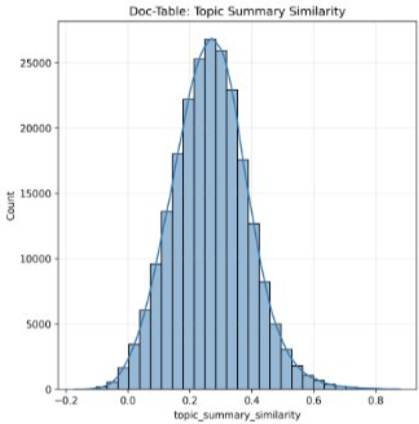}
\caption{\small Left: topic-title similarity distribution. Right: topic-summary similarity distribution. Both are bell-shaped.}
\label{fig:doc_tab_topic_title}
\label{fig:doc_tab_topic_summary}
\end{figure}

\paragraph{Topic-title and topic-summary similarity}
Figure~\ref{fig:doc_tab_topic_title} show the distributions of topic-title similarity and topic-summary similarity, respectively. Both curves are bell-shaped, indicating most document-table pairs are moderately similar, with few extreme cases. These two histograms are placed side by side for direct comparison.

\begin{figure}[H]
\centering
\includegraphics[width=0.48\linewidth]{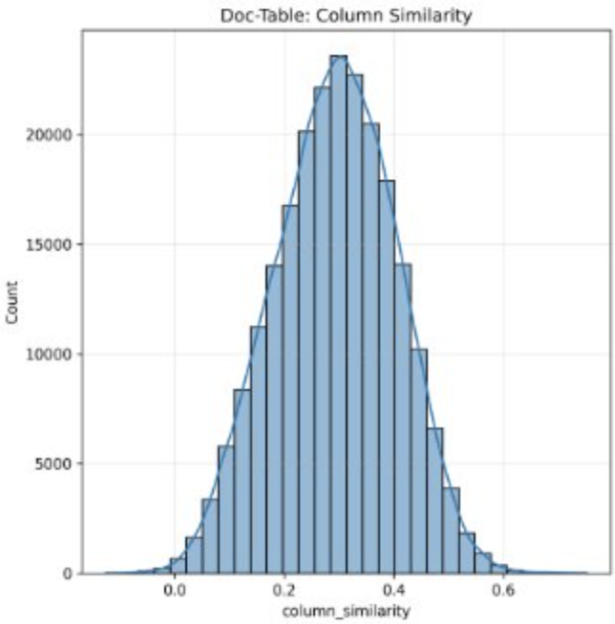}
\includegraphics[width=0.48\linewidth]{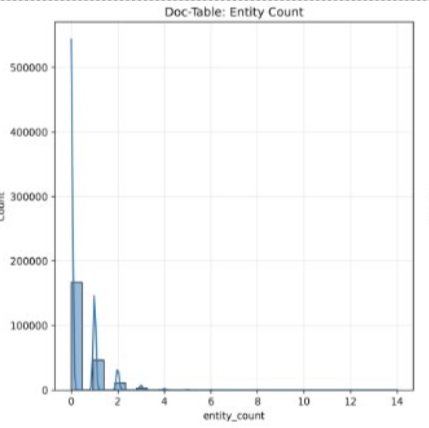}
\caption{\small Left: column similarity distribution. Right: entity count per table chunk (left-skewed).}
\label{fig:doc_tab_column}
\label{fig:doc_tab_entity}
\end{figure}

\paragraph{Column similarity and entity counts}
Figure~\ref{fig:doc_tab_column} shows column similarity and number of entities per table chunk. Column similarity exhibits a bell-shaped distribution, while the entity count is left-skewed, consistent with document-document observations. Placing these side by side highlights the distributional patterns across structural and semantic features.

\begin{figure}[H]
\centering
\includegraphics[width=0.32\linewidth]{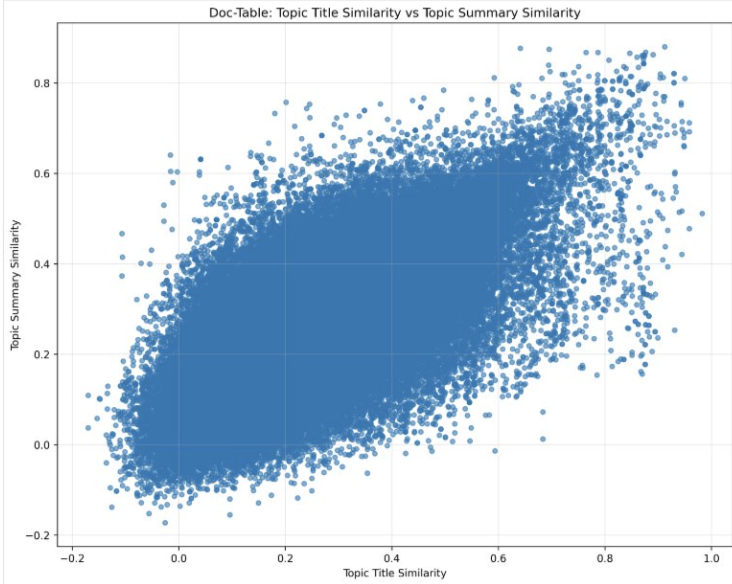}
\includegraphics[width=0.32\linewidth]{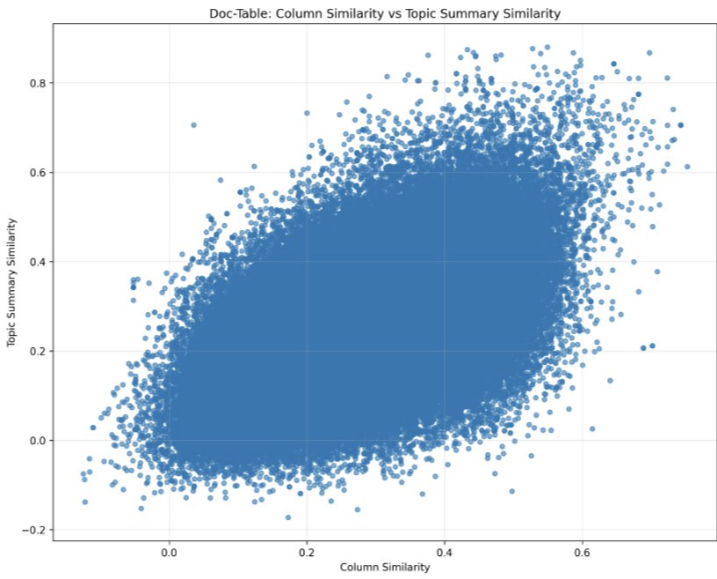}
\includegraphics[width=0.32\linewidth]{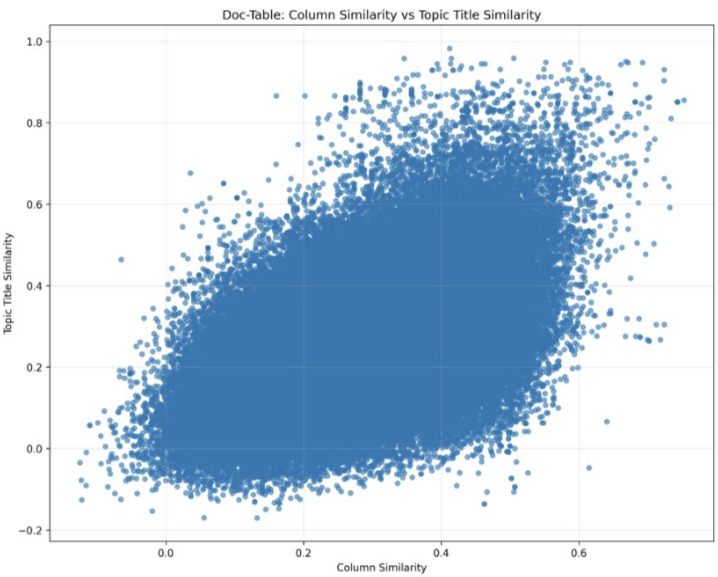}
\caption{\small Scatter plots showing alignment between document and table chunks: topic-title vs topic-summary (left), column vs topic-summary (center), column vs topic-title (right). Diagonal trends indicate strong alignment.}
\label{fig:doc_tab_topic_title_summary}
\label{fig:doc_tab_column_topic_summary}
\label{fig:doc_tab_column_topic_title}
\end{figure}

\paragraph{Scatter plots of topic, title, summary, and column similarities}
Figure~\ref{fig:doc_tab_topic_title_summary} shows scatter plots for combinations of topic, title, summary, and column similarities. Diagonal trends indicate strong alignment between document and table chunks across multiple semantic dimensions.

\paragraph{Correlation matrix}
Table~\ref{tab:doc_tab_matrix} shows the correlation matrix of document-table similarity metrics. Column similarity, title similarity, and summary similarity are moderately correlated, confirming that cross-modal semantic alignment can be captured using simple embeddings without LLM-based processing.

\begin{table}[H]
    \centering
    \small
    \begin{tabular}{lcccc}
        \toprule
        & \textbf{(1)} & \textbf{(2)} & \textbf{(3)} & \textbf{(4)} \\
        \midrule
        \textbf{(1) Col Sim}     & 1.000 & 0.078 & 0.517 & 0.535 \\
        \textbf{(2) Ent Count}   & 0.078 & 1.000 & 0.107 & 0.107 \\
        \textbf{(3) Title Sim}   & 0.517 & 0.107 & 1.000 & 0.691 \\
        \textbf{(4) Summ Sim}    & 0.535 & 0.107 & 0.691 & 1.000 \\
        \bottomrule
    \end{tabular}
    \caption{Doc-Table Correlation Matrix. Abbreviations: (1) column\_similarity, (2) entity\_count, (3) topic\_title\_similarity, and (4) topic\_summary\_similarity}
    \label{tab:doc_tab_matrix}
\end{table}

\section{Runtime Analysis of all State of the Art baselines for STaRK}\label{sec:runtime}

\subsection{4StepFocus}

4StepFocus processes queries over semi-structured knowledge bases (SKBs) through symbolic prefiltering, followed by vector similarity search and LLM-based reranking.

\textbf{Pipeline:} 
\begin{enumerate}
    \item \textbf{LLM Extraction:} The LLM extracts triplets $T$ and the target variable $x_{\text{target}}$ from the query $q$.
    \item \textbf{Symbolic Prefiltering (SUBSTITUTE):} Iteratively intersects candidate sets with KG neighbors until convergence, producing a filtered candidate set $C_{\text{filtered}}$.
    \item \textbf{Vector Similarity Scoring (VSS):} Scores $C_{\text{filtered}}$ using unstructured data, returning the top-$k_{\max}$ candidates along with additional relevant candidates.
    \item \textbf{LLM Reranking:} Reranks candidates using the SKB context.
\end{enumerate}

\textbf{SUBSTITUTE Algorithm:}

\begin{algorithm}[h]
\SetAlgoLined
\DontPrintSemicolon
\KwIn{Candidate set $C$, triplets $T = \{(h_i, e_i, t_i)\}$, target variable $x_{\text{target}}$}
\KwOut{Filtered candidate set $C_{\text{filtered}}$}

\ForEach{variable $x$}{
    $\text{substitute}[x] \gets \{ c \in C \mid \text{type}(c) = \text{type}(x) \}$\;
}
\Repeat{no changes in any $\text{substitute}[x]$}{
    \ForEach{$(h_i, e_i, t_i) \in T$}{
        $\text{substitute}[h_i] \gets \text{substitute}[h_i] \cap \text{neighbors}(h_i, e_i, -)$\;
        $\text{substitute}[t_i] \gets \text{substitute}[t_i] \cap \text{neighbors}(t_i, e_i, +)$\;
    }
}
$C_{\text{filtered}} \gets \text{substitute}[x_{\text{target}}]$\;
\Return $C_{\text{filtered}}$\;
\caption{SUBSTITUTE: Symbolic Prefiltering for Candidate Selection}
\end{algorithm}

\textbf{Runtime Analysis:} 
\begin{itemize}
    \item \textbf{Step 1:} $O(1)$ - single LLM call.
    \item \textbf{Step 2:} $O(n \cdot |T| \cdot \deg)$ - $n$ iterations over $|T|$ triplets with average node degree $\deg$.
    \item \textbf{Step 3:} $O(|C_{\text{filtered}}|)$ - VSS applied to filtered candidates.
    \item \textbf{Step 4:} $O(1)$ - single LLM call for reranking.
\end{itemize}

\textbf{Total Complexity:} 
\[
O(n \cdot |T| \cdot \deg + |C_{\text{filtered}}|)
\]

\subsection{FocusedR}

FocusedR combines Cypher query generation, symbolic grounding, and hybrid retrieval (Cypher-constrained plus fallback Vector similarity search(VSS)) before LLM-based reranking.

\textbf{Pipeline:} 
\begin{enumerate}
    \item \textbf{LLM Cypher Generation:} The LLM generates a Cypher query $q_{\text{cypher}}$ from $q$ using node and edge types.
    \item \textbf{Regex Parsing:} Parse $q_{\text{cypher}}$ into target $y$, triplets $T$, and symbol constraints $S_{\text{raw}}$.
    \item \textbf{Symbolic Grounding:} 
    \begin{itemize}
        \item \textbf{SYMBOL\_CANDIDATES:} Perform VSS for symbol constants.
        \item \textbf{GROUND\_TRIPLETS:} Iteratively prune candidate sets via neighbor intersections.
    \end{itemize}
    \item \textbf{Adaptive Candidate Expansion:} While $|C_{\text{cypher}}| < k$, exponentially increase $\hat{k}$ and repeat Step 3.
    \item \textbf{Hybrid VSS Retrieval:} Perform two VSS calls:
    \begin{itemize}
        \item $Y_{\text{cypher}}$: Cypher-constrained candidates
        \item $Y_{\text{vss}}$: Fallback unstructured candidates
    \end{itemize}
    \item \textbf{LLM Reranking:} Rerank concatenated candidates $Y = Y_{\text{cypher}} + Y_{\text{vss}}$.
\end{enumerate}

\textbf{GROUND\_TRIPLETS Algorithm:}

\begin{algorithm}[h]
\SetAlgoLined
\DontPrintSemicolon
\KwIn{Triplets $T = \{(h_i, e_i, t_i)\}$, symbol candidate sets $S$}
\KwOut{Pruned symbol candidate sets $S$}

\Repeat{no changes in any $S(x)$}{
    \ForEach{$(h_i, e_i, t_i) \in T$}{
        $S(h_i) \gets S(h_i) \cap \text{neighbors}(h_i, e_i, -)$\;
        $S(t_i) \gets S(t_i) \cap \text{neighbors}(t_i, e_i, +)$\;
    }
}
\Return{$S$}\;
\caption{GROUND\_TRIPLETS: Iterative Candidate Pruning}
\end{algorithm}

\textbf{Runtime Analysis:} 
\begin{itemize}
    \item \textbf{Step 1:} $O(1)$ - single LLM call.
    \item \textbf{Steps 2--4:} $O(n_1 \cdot n_2 \cdot |T| \cdot \deg)$ - $n_1$ outer $\hat{k}$ expansions, $n_2$ inner convergence iterations, $|T|$ triplets, $\deg$ average degree.
    \item \textbf{Step 5:} $O(|C_{\text{cypher}}| + |C(y'.\text{type})|)$ - two VSS calls.
    \item \textbf{Step 6:} $O(1)$ - single LLM call for reranking.
\end{itemize}

\textbf{Total Complexity:} 
\[
O(n_1 \cdot n_2 \cdot |T| \cdot \deg + |C_{\text{cypher}}|)
\]

\subsection{AvaTaR}

AvaTaR is an agent optimization framework that improves tool-calling LLMs through contrastive reasoning. It is typically instantiated as a ReAct-style agent operating over retrieval tools.

\textbf{Pipeline:} 
\begin{enumerate}
    \item \textbf{Agent Initialization:} Initialize the agent with query $q$ and a toolset (e.g., retrieval APIs over an SKB).
    \item \textbf{Agent Loop:} Repeat for $m$ steps:
    \begin{itemize}
        \item The LLM generates a \emph{thought} and an \emph{action} (tool call).
        \item Execute the action and append the resulting observation to the trajectory.
    \end{itemize}
    \item \textbf{Termination:} Continue until maximum steps, success condition, or a final answer is generated.
    \item \textbf{AvaTaR Optimization (Training-time):} Perform contrastive fine-tuning using successful vs.\ failed trajectories to improve future performance.
\end{enumerate}

\textbf{Runtime Analysis:} 
\begin{itemize}
    \item Per trajectory: $O(m)$ LLM calls and $O(m_r)$ retrievals ($m$ = total steps, $m_r \le m$ = number of retrieval actions).  
    \item No explicit set operations; reasoning is implicit in the LLM.  
    \item No dedicated reranker; retrieval results are fed directly to the agent prompt.  
    \item Inference can be amortized over multiple queries via the trained policy.
\end{itemize}

\textbf{Total Complexity (per query):}  
\[
O(m) \text{ LLM calls} + O(m_r) \text{ retrievals}
\]
where $m$ is the trajectory length until stopping condition.

\subsection{KAR}

KAR performs knowledge-aware query expansion through entity extraction, neighbor retrieval, document relation filtering, triple construction, and final retrieval.

\textbf{Pipeline:} 
\begin{enumerate}
    \item \textbf{Entity Extraction:} The LLM extracts entities $E = \{e_1, \dots, e_e\}$ from query $q$.
    \item \textbf{Neighbor and Document Retrieval:} For each entity $e_i$, retrieve neighbors $N(e_i)$ and associated documents $D_i$ (total $e+1$ retrievals).
    \item \textbf{Document Relation Filtering:} Filter retrieved documents based on cosine similarity over the $e+1$ neighbor sets.
    \item \textbf{Triple Construction:} Construct triples $(e_i, r_j, d_k)$ from the filtered neighbor sets.
    \item \textbf{Query Expansion:} Feed all triples to the LLM to generate an expanded query $q_{\text{exp}}$.
    \item \textbf{Final Retrieval:} Execute a single retrieval over the expanded query $q_{\text{exp}}$.
\end{enumerate}

\textbf{Runtime Analysis:} 
\begin{itemize}
    \item Step 1: $O(1)$ - single LLM call for entity extraction.
    \item Step 2: $O(e)$ - $e$ entity neighbor retrievals plus 1 document retrieval.
    \item Step 3: $O((e+1)\cdot \deg_D)$ - cosine similarity over neighbor documents, $\deg_D$ = average doc neighbors.
    \item Step 4: $O(|T|)$ - triple construction, where $|T|$ is total triples.
    \item Step 5: $O(1)$ - single LLM call for query expansion.
    \item Step 6: $O(1)$ - single final retrieval.
\end{itemize}

\textbf{Total Complexity:} 
\[
O(e \cdot \deg_D + |T|)
\]
with exactly two LLM calls.

\subsection{ReAct}

ReAct interleaves LLM reasoning with tool calls (typically retrieval) in a think-act-observe loop until task completion or a step limit.

\textbf{Pipeline:} 
\begin{enumerate}
    \item \textbf{Initial LLM Reasoning:} The LLM generates an initial thought and first action (retrieval tool call).
    \item \textbf{Execute Action:} Execute the retrieval action and append results as an observation to the trajectory.
    \item \textbf{Subsequent Reasoning:} LLM reasons over the observation and generates the next thought and action.
    \item \textbf{Iteration:} Repeat Steps 2--3 for $m$ iterations until a final answer is obtained or maximum steps are reached.
\end{enumerate}

\textbf{Runtime Analysis:} 
\begin{itemize}
    \item Per trajectory: $O(m)$ LLM calls and $O(m_r)$ retrievals.
    \item $m$: total steps until stopping condition (max steps, success, or final answer).  
    \item $m_r \le m$: number of retrieval actions (subset of total actions).  
    \item No explicit set operations; reasoning is implicit in the LLM.  
    \item No dedicated reranker; retrievals are fed directly to the agent prompt.
\end{itemize}

\textbf{Total Complexity (per query):}  
\[
O(m) \text{ LLM calls} + O(m_r) \text{ retrievals}
\]
where $m$ is the trajectory length until stopping.

\subsection{Reflexion}
Reflexion extends ReAct agents with self-reflection: after each failed episode, an LLM analyzes the trajectory to generate verbal feedback for the next episode.

\textbf{Pipeline:} 
\begin{itemize}
\item \textbf{Step 1:} Run ReAct episode: $m$ iterations of LLM call $\rightarrow$ retrieval $\rightarrow$ observation.
\item \textbf{Step 2:} Evaluate episode outcome (success/failure).
\item \textbf{Step 3:} If failed, LLM generates reflection $r_i$ analyzing trajectory and errors.
\item \textbf{Step 4:} Append reflection to prompt, repeat Steps 1--3 for $E$ episodes total.
\end{itemize}

\textbf{Runtime analysis:} 
\begin{itemize}
\item Per episode: $O(m)$ LLM calls + $O(m_r)$ retrievals ($m$: steps, $m_r$: retrievals).
\item Per reflection: $O(1)$ LLM call (self-critique).
\item $E$: total episodes until success or cap.
\item $R$: reflections ($R \leq E-1$, one per failed episode).
\item Total: $O(E\cdot m + R)$ LLM calls + $O(E\cdot m_r)$ retrievals.
\end{itemize}

\textbf{Total:} $O(E\cdot m + R)$ LLM calls + $O(E\cdot m_r)$ retrievals, where $E$ is episodes, $m$ is steps per episode, $R$ is reflections.

\subsection{\methodname using \agentname}

\methodname using \agentname queries through a single-pass agentic retrieval pipeline with hybrid dense+sparse expansion.

\textbf{Pipeline:} 
\begin{enumerate}
    \item \textbf{Query Generation:} Agent LLM generates a Cypher/HNSW query from the natural language input $q$.
    \item \textbf{Initial Retrieval:} Retrieve the top-$k$ nodes $N_k$ using the generated query.
    \item \textbf{Graph Expansion:} Retrieve 1-hop neighbors $N(N_k)$ of the top-$k$ nodes.
    \item \textbf{Hybrid Retrieval:} Apply dense+sparse retrieval over $N(N_k)$ to produce the final $k + k'$ candidate set.
\end{enumerate}

\textbf{Runtime Analysis:} 
\begin{itemize}
    \item Step 1: $O(1)$ - single agent LLM call.
    \item Step 2: $O(k)$ - initial retrieval (HNSW/Cypher).
    \item Step 3: $O(k \cdot \deg)$ - neighbor expansion ($\deg$ = average node degree).
    \item Step 4: $O(k \cdot \deg)$ - hybrid dense+sparse retrieval over neighbors.
\end{itemize}

\textbf{Total Complexity:} 
\[
O(k \cdot \deg)
\]
with exactly one LLM call and no iterative loops.

\section{Effect of Dense+Sparse Retriever + Graph Expansion on STaRK Dataset}\label{sec:stark_k}

\subsection{AMAZON}
\begin{figure}[H]
\centering
\includegraphics[width=0.95\linewidth]{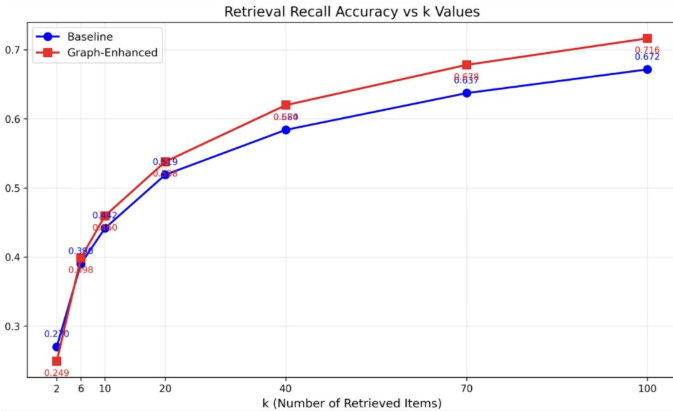}
\caption{\small Recall@k for Synthetic queries on AMAZON. Graph expansion improves recall across all $k$.}
\label{fig:amazon_syn}
\end{figure}

\begin{figure}[H]
\centering
\includegraphics[width=1.0\linewidth]{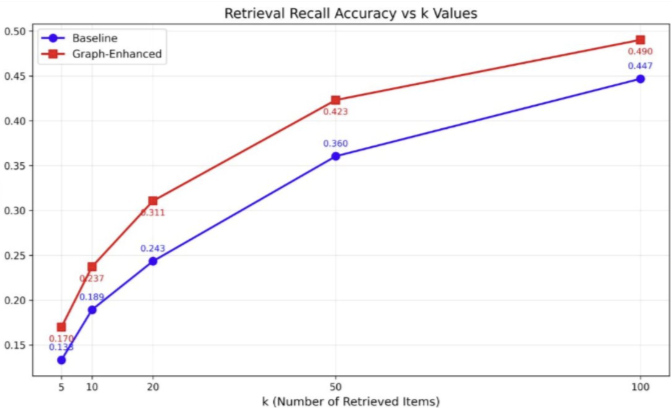}
\caption{\small Recall@k for Human queries on AMAZON. Graph expansion consistently increases recall.}
\label{fig:amazon_hum}
\end{figure}

Figures~\ref{fig:amazon_syn} and~\ref{fig:amazon_hum} show recall@k for Synthetic and Human queries on the AMAZON subset. Graph expansion consistently improves recall across all $k$ values. This aligns with our observation that nodes in AMAZON contain rich product descriptions and feature-rich metadata, providing ample semantic content for propagation. The one-hop expansion strategy reliably increases coverage of relevant nodes without introducing noise, yielding a consistent gain over the Dense+Sparse baseline.

\subsection{MAG}
\begin{figure}[H]
\centering
\includegraphics[width=1.0\linewidth]{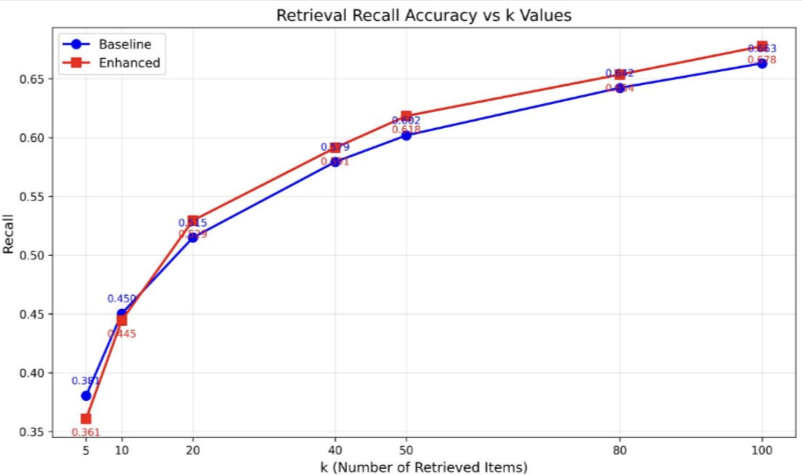}
\caption{\small Recall@k for Synthetic queries on MAG. Graph expansion improves recall at higher $k$, but hurts at low $k$.}
\label{fig:mag_syn}
\end{figure}

\begin{figure}[H]
\centering
\includegraphics[width=1.0\linewidth]{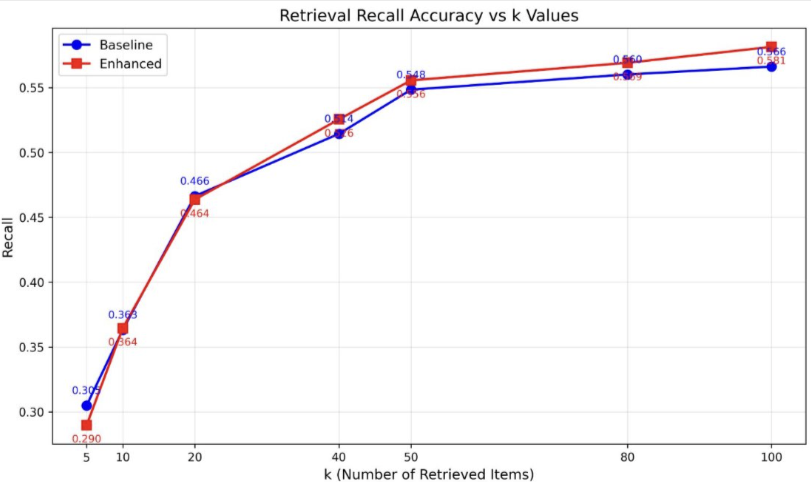}
\caption{\small Recall@k for Human queries on MAG. Graph expansion shows gains only at larger $k$.}
\label{fig:mag_hum}
\end{figure}

Figures~\ref{fig:mag_syn} and~\ref{fig:mag_hum} show recall@k for MAG. Here, graph expansion helps primarily at higher $k$ values ($k\gtrsim 10$), while recall may drop at lower $k$. MAG nodes contain shorter titles and abstracts, so expanding the graph at small $k$ can introduce weakly relevant neighbors, hurting early-rank retrieval. At larger $k$, however, graph augmentation still provides additional relevant nodes, boosting recall and demonstrating its utility when more candidates are available.

\subsection{PRIME}
\begin{figure}[H]
\centering
\includegraphics[width=1.0\linewidth]{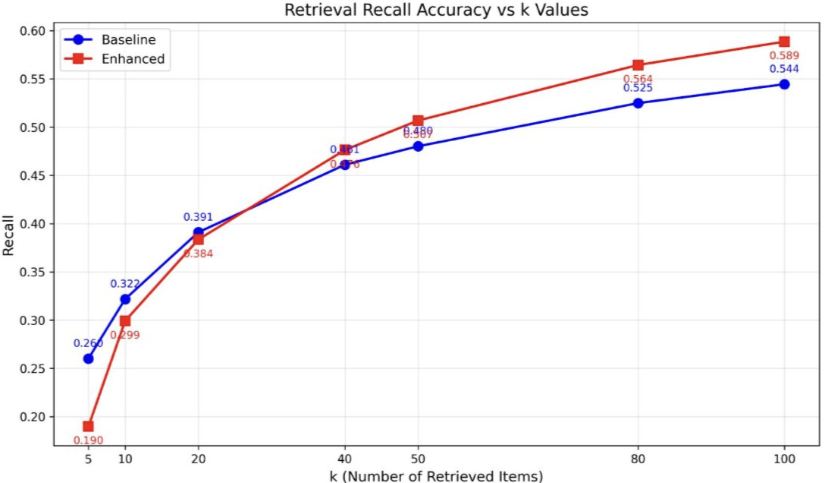}
\caption{\small Recall@k for Synthetic queries on PRIME. Graph expansion is effective only at higher $k$.}
\label{fig:prime_syn}
\end{figure}

\begin{figure}[H]
\centering
\includegraphics[width=1.0\linewidth]{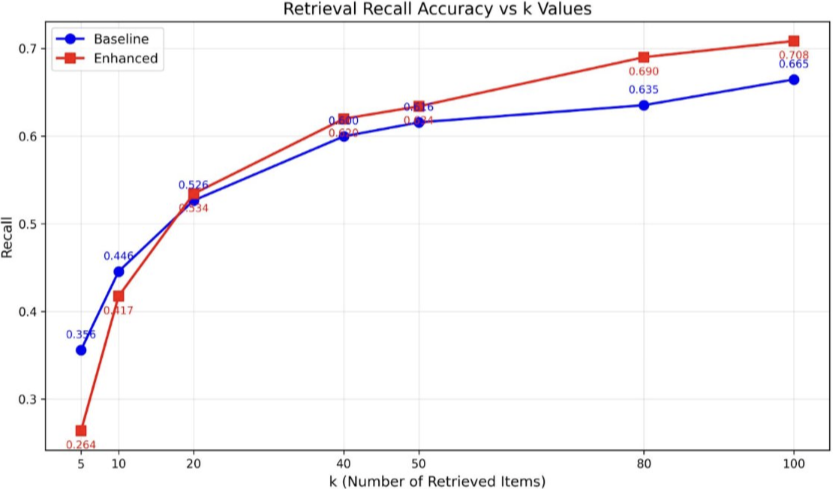}
\caption{\small Recall@k for Human queries on PRIME. Graph expansion helps mainly at larger $k$.}
\label{fig:prime_hum}
\end{figure}

Figures~\ref{fig:prime_syn} and~\ref{fig:prime_hum} show recall@k for PRIME. Similar to MAG, graph expansion benefits are observed primarily at higher $k$ values ($k \gtrsim 20$), while lower $k$ sees slight drops in recall. PRIME nodes contain concise drug, pathway, and disease descriptions, limiting semantic propagation. These trends reinforce our finding that graph-based expansion is most effective when nodes have rich textual content.

\section{Implementation Details.}
Our system follows a two-stage design with offline preprocessing and online query processing. Offline, we perform semantic chunking and graph construction using the all-MiniLM-L6-v2 embedding model (384-d),  \cite{wang2020minilm} and build ANN indices with HNSW (M=64); preprocessing is executed once on a single NVIDIA A100 GPU. For explicit schema graphs, we store and query the structured data in Neo4j, and extract lightweight metadata such as titles and entities from nodes/fields to support linking and retrieval. Online, we answer each query by combining sparse and dense signals via linear fusion (\(\alpha=0.6\) for BM25, \(\beta=0.4\) for embeddings), and we use GPT-4o-mini-2024-07-18 \cite{gpt4ocard} where LLM-based extraction/planning is required.

\section{Additional Metrics for STaRK}

Table~\ref{tab:hmetrics_compact} presents Hits@1, Hits@5, and MRR scores for all methods across AMAZON, MAG, and PRIME datasets.

\begin{table*}[t]
\small
\centering
\setlength{\tabcolsep}{1.25pt}
\begin{tabular}{l|ccc|ccc|ccc|ccc|ccc|ccc}
\toprule
 & \multicolumn{6}{c|}{Amazon} & \multicolumn{6}{c|}{MAG} & \multicolumn{6}{c}{Prime} \\
Method
 & \multicolumn{3}{c}{Human} & \multicolumn{3}{c|}{Synthetic}
 & \multicolumn{3}{c}{Human.} & \multicolumn{3}{c|}{Synthetic}
 & \multicolumn{3}{c}{Human.} & \multicolumn{3}{c}{Synthetic} \\
 & H1 & H5 & M & H1 & H5 & M
 & H1 & H5 & M & H1 & H5 & M
 & H1 & H5 & M & H1 & H5 & M \\
\hline\noalign{\vskip 1pt}
\multicolumn{19}{c}{\textbf{Dense}} \\
\hline\noalign{\vskip 1pt}
ada-002 & 39.50 & 64.19 & 52.65 & 29.08 & 49.61 & 38.62
        & 28.57 & 41.67 & 35.81 & 29.08 & 49.61 & 38.62
        & 17.35 & 34.69 & 26.35 & 12.63 & 31.49 & 21.41 \\
m-ada & 46.91 & 72.84 & 58.74 & 40.07 & 64.98 & 51.55
      & 23.81 & 41.67 & 31.43 & 25.92 & 50.43 & 36.94
      & 24.49 & 39.80 & 32.98 & 15.10 & 33.56 & 23.49 \\
DPR & 16.05 & 39.51 & 27.21 & 15.29 & 47.93 & 30.20
    & 4.72 & 9.52 & 7.90 & 10.51 & 35.23 & 21.34
    & 2.04 & 9.18 & 7.05 & 4.46 & 21.85 & 12.38 \\
\hline\noalign{\vskip 1pt}
\multicolumn{19}{c}{\textbf{Sparse}} \\
\hline\noalign{\vskip 1pt}
BM25 & 27.16 & 51.85 & 18.79 & 25.85 & 45.25 & 34.91
     & 32.14 & 41.67 & 37.42 & 25.85 & 45.25 & 34.91
     & 22.45 & 41.84 & 30.37 & 12.75 & 27.92 & 19.84 \\
\hline\noalign{\vskip 1pt}
\multicolumn{19}{c}{\textbf{Structural}} \\
\hline\noalign{\vskip 1pt}
QAGNN & 22.22 & 49.38 & 31.33 & 12.88 & 39.01 & 29.12
      & 20.24 & 26.19 & 25.53 & 12.88 & 39.01 & 29.12
      & 6.12 & 13.27 & 26.35 & 8.85 & 21.35 & 14.73 \\
\hline\noalign{\vskip 1pt}
\multicolumn{19}{c}{\textbf{Hybrid}} \\
\hline\noalign{\vskip 1pt}
4Step & -- & -- & -- & 47.60 & 67.60 & 56.50
      & -- & -- & -- & 53.80 & 69.20 & 61.40
      & -- & -- & -- & 39.30 & 53.20 & 45.80 \\
FocusR & -- & -- & -- & 64.00 & 76.20 & 69.30
       & -- & -- & -- & 74.10 & 84.20 & 78.80
       & -- & -- & -- & 46.40 & 63.90 & 53.70 \\
D+B & 32.14 & 44.05 & 38.46 & 29.72 & 49.64 & 38.71
    & 32.14 & 44.05 & 38.46 & 29.72 & 49.64 & 38.71
    & 24.77 & 43.12 & 33.25 & 13.71 & 31.13 & 21.39 \\
\hline\noalign{\vskip 1pt}
\multicolumn{19}{c}{\textbf{Query Expansion}} \\
\hline\noalign{\vskip 1pt}
HyDe & 45.68 & 72.84 & 57.56 & 28.98 & 50.10 & 39.58
     & 33.33 & 44.05 & 38.95 & 28.98 & 50.10 & 39.58
     & 24.77 & 42.20 & 33.65 & 16.85 & 37.59 & 26.56 \\
RAR & 55.56 & 71.60 & 62.15 & 39.02 & 52.87 & 39.58
    & 38.10 & 45.24 & 42.04 & 39.02 & 52.87 & 39.58
    & 31.19 & 43.12 & 37.72 & 22.23 & 40.84 & 30.93 \\
AGR & 55.56 & 71.60 & 63.54 & 39.29 & 53.66 & 46.20
    & 33.33 & 44.05 & 38.95 & 39.29 & 53.66 & 46.20
    & 32.11 & 49.54 & 39.27 & 25.85 & 44.41 & 35.04 \\
KAR & 61.73 & 72.84 & 66.32 & 50.47 & 65.37 & 57.51
    & 51.20 & 58.30 & 54.20 & 50.47 & 65.37 & 57.51
    & 44.95 & 60.55 & 44.51 & 30.35 & 49.30 & 39.22 \\
\hline\noalign{\vskip 1pt}
\multicolumn{19}{c}{\textbf{Agentic}} \\
\hline\noalign{\vskip 1pt}
AvaTaR & 58.32 & 76.54 & 65.91 & 49.97 & 69.16 & 52.01
       & 33.33 & 42.86 & 38.62 & 46.08 & 59.32 & 52.01
       & 33.03 & 51.37 & 41.00 & 20.10 & 39.89 & 29.18 \\
ReACT & 45.65 & 71.73 & 58.81 & 42.14 & 64.56 & 52.30
      & 27.27 & 40.00 & 33.94 & 31.07 & 49.49 & 39.25
      & 21.73 & 33.33 & 28.20 & 15.28 & 31.95 & 22.76 \\
Reflex & 49.38 & 64.19 & 52.91 & 42.79 & 65.05 & 52.91
       & 28.57 & 39.29 & 36.53 & 40.71 & 54.44 & 47.06
       & 16.52 & 33.03 & 23.99 & 14.28 & 34.99 & 24.82 \\
\hline\noalign{\vskip 1pt}
\multicolumn{19}{c}{\textbf{Finetuned}} \\
\hline\noalign{\vskip 1pt}
mFAR & -- & -- & -- & 41.20 & 70.00 & 54.20
     & -- & -- & -- & 49.00 & 69.60 & 58.20
     & -- & -- & -- & 40.90 & 62.80 & 51.20 \\
MoR & -- & -- & -- & 52.20 & 74.60 & 62.20
    & -- & -- & -- & 58.20 & 78.30 & 67.10
    & -- & -- & -- & 36.40 & 60.00 & 46.90 \\
\hline\noalign{\vskip 1pt}
\multicolumn{19}{c}{\textbf{Ours}} \\
\hline\noalign{\vskip 1pt}
\agentname & 22.22 & 58.03 & 39.93 & 26.72 & 54.27 & 36.24
      & 44.40 & 54.80 & 49.40 & 41.00 & 56.41 & 49.30
      & 41.66 & 61.14 & 52.53 & 29.12 & 48.31 & 41.94 \\
A+G+E & 25.18 & 55.13 & 37.75 & 25.18 & 53.45 & 36.18
      & 45.10 & 55.60 & 48.90 & 41.70 & 57.20 & 48.80
      & 42.20 & 61.60 & 51.53 & 30.14 & 49.56 & 42.79 \\
A+G-E & 28.41 & 63.84 & 42.48 & 28.41 & 55.13 & 42.48
    & 43.20 & 58.70 & 49.40 & 43.20 & 60.00 & 49.30
    & 40.23 & 65.40 & 52.53 & 30.86 & 51.48 & 43.20 \\
\bottomrule
\end{tabular}
\caption{Hits@1 (H@1), Hits@5 (H@5), and Mean Reciprocal Rank (MRR) scores on STaRK datasets. A = \agentname, G=\methodname, E=Edge}
\label{tab:hmetrics_compact}
\end{table*}

\newpage
\section{Prompt for LLM Metadata Extraction}
\subsection{Entity Extraction for Table Chunks}
\begin{lstlisting}[language=Prompt,caption={Table metadata extraction prompt}]

You are a table metadata extraction engine. Given table data with headers and rows, parse it into a single valid JSON object following exactly this schema:

{{
  "table_title": "<verbatim table title or inferred title if unavailable>",
  "table_description": "<one-sentence explanation of the table content>",
  "table_summary": "<concise interpretation or insight derived from the data>",
  "col_desc": {{
    "<ColumnName1>": "<contextual description of the column's representation and purpose>",
    ...
  }},
  "col_format": {{
    "<ColumnName1>": "<data format specification: string, number, date (format), etc.>",
    ...
  }},
  "entities": {{
    "<EntityName1>": {{
      "type": "<person/place/organization/event/concept/thing>",
      "category": "<named/non-named>",
      "description": "<brief description of the entity>"
    }},
    ...
  }}
}}

Instructions:
1. table_title: Use the source name if available; otherwise, infer a descriptive title from the data.
2. table_description: Provide one sentence explaining the information contained in this table.
3. table_summary: Provide a brief insight or interpretation of the overall data patterns.
4. col_desc: For each column header, provide a contextual description of what the column represents, its purpose, and significance.
5. col_format: For each column header, specify the data format (e.g., string - names, number - scores, date (Month Day)).
6. entities: Extract all unique entities mentioned in the table with detailed information:
   - type: person, place, organization, event, concept, thing, etc.
   - category: named (proper nouns) or non-named (common nouns/concepts)
   - description: brief explanation of what this entity represents

Output: Pure JSON only. No markdown or comments.

One-Shot Example:

Table Source Name: "1911 Notre Dame Fighting Irish football team - Schedule"
Table Data:
[
  {{
    "Date": "October 7",
    "Opponent": "Ohio Northern",
    "Site": "Cartier Field South Bend , IN",
    "Result": "W 32-6"
  }},
  {{
    "Date": "October 14", 
    "Opponent": "St. Viator",
    "Site": "Cartier Field South Bend , IN",
    "Result": "W 43-0"
  }},
  {{
    "Date": "October 21",
    "Opponent": "Butler", 
    "Site": "Cartier Field South Bend , IN",
    "Result": "W 27-0"
  }}
]

Expected Output:
{{
  "table_title": "1911 Notre Dame Fighting Irish football team - Schedule",
  "table_description": "This table presents the football schedule and results for the 1911 Notre Dame Fighting Irish football team.",
  "table_summary": "Notre Dame had a strong season with multiple wins at home field, scoring consistently high points against various opponents.",
  "col_desc": {{
    "Date": "The scheduled date when each football game was played during the 1911 season",
    "Opponent": "The opposing football team that Notre Dame played against in each game",
    "Site": "The venue and location where each game took place, including home and away games",
    "Result": "The final outcome and score of each game showing Notre Dame's performance"
  }},
  "col_format": {{
    "Date": "string - date (Month Day format)",
    "Opponent": "string - team names",
    "Site": "string - locations with venue details",
    "Result": "string - game results (W/T/L followed by score)"
  }},
  "entities": {{
    "Notre Dame Fighting Irish": {{
      "type": "organization",
      "category": "named",
      "description": "College football team from the University of Notre Dame"
    }},
    "Ohio Northern": {{
      "type": "organization", 
      "category": "named",
      "description": "Opposing college football team"
    }},
    "St. Viator": {{
      "type": "organization",
      "category": "named", 
      "description": "Opposing college football team"
    }},
    "Butler": {{
      "type": "organization",
      "category": "named",
      "description": "Opposing college football team"
    }},
    "Cartier Field": {{
      "type": "place",
      "category": "named",
      "description": "Notre Dame's home football stadium"
    }},
    "South Bend": {{
      "type": "place",
      "category": "named",
      "description": "City in Indiana where Notre Dame is located"
    }},
    "IN": {{
      "type": "place",
      "category": "named",
      "description": "Indiana state abbreviation"
    }}
  }}
}}

Now, analyze this table data:

Table Source Name: "{source_name}"
Table Data:
{table_content}

\end{lstlisting}
\subsection{Entity Extraction for Document Chunks}
\begin{lstlisting}[language=Prompt,caption={Document metadata extraction prompt}]
You are a metadata extraction engine. Given a block of text, parse it into a single valid JSON object following exactly this schema:

{
  "entities": {
    "<EntityName1>": {
      "details": [
        "<detailed attribute or description sentence>",
        …
      ]
    },
    …
  },
  "events": {
    "<EventName1>": {
      "date": "<YYYY-MM-DD or null>",
      "details": "<one-sentence context or significance>"
    },
    …
  },
  "timeline": [
    "<ISO date - description>",
    …
  ],
  "topic": "<one-sentence description of the overall content>"
}

Instructions:
1. entities: One key per unique NAMED entity (not a noun phrase). Nicknames and abbreviations count as separate entities, but the original must be mentioned in details (e.g., ORIGINAL: ...).
   - details: Distinguishing descriptive sentences or attributes where the entity is the subject.
2. events: One key per named event.
   - date: ISO date or null.
   - details: A single-sentence summary of its significance.
3. timeline: Chronological array of "<date> - <brief description>" for all dated mentions.
4. topic: One sentence summarizing the overall theme.

Output: Pure JSON only. No markdown or comments.
Note: Each object must have unique keys.

One‐Shot Example

Input Text

Bixente Jean Michel Lizarazu (Basque pronunciation: [biˈʃente liˈs̪araˌs̪u], born 9 December 1969) is a French former professional footballer who played as a left-back.

He rose through the ranks at Bordeaux and finished second in the French First Division in 1989-1990. The team was relegated but won promotion from Second Division in 1991-92. His Bordeaux team finished runners-up in the 1995-96 UEFA Cup.

In 1997, he joined Bayern Munich and won six Bundesliga championships and the 2000-01 UEFA Champions League, scoring in the final shootout.

Expected Output
{
  "entities": {
    "Bixente Jean Michel Lizarazu": {
      "details": [
        "Basque pronunciation: [biˈʃente liˈs̪araˌs̪u]",
        "Former professional footballer",
        "Represented France at international level",
        "Born 9 December 1969",
        "Played as a left-back"
      ]
    },
    "Bordeaux FC": {
      "details": [
        "French football club",
        "Club where Lizarazu began his professional career"
      ]
    },
    "Bayern Munich": {
      "details": [
        "German club Lizarazu joined in 1997",
        "Top-tier Bundesliga club"
      ]
    },
    "Bundesliga": {
      "details": [
        "Top division of German football"
      ]
    },
    "France": {
      "details": [
        "Country where Lizarazu was born and began his football career"
      ]
    },
    "French First Division": {
      "details": [
        "Top-tier football league in France (now Ligue 1)"
      ]
    },
    "French Second Division": {
      "details": [
        "Second-tier football league in France (now Ligue 2)"
      ]
    }
  },
  "events": {
    "1989-1990 French First Division season": {
      "date": "1989-1990",
      "details": "Early top-tier success in France"
    },
    "1991-1992 French Second Division promotion": {
      "date": "1991-1992",
      "details": "Return to French First Division after relegation"
    },
    "1995-1996 UEFA Cup Final": {
      "date": "1996-05-08",
      "details": "Major European final for a French club"
    },
    "1997 Transfer to Bayern Munich": {
      "date": "1997-07-01",
      "details": "Transfer from French to German football"
    },
    "2001 UEFA Champions League Final": {
      "date": "2001-05-23",
      "details": "Lizarazu secured European title with German club"
    }
  },
  "timeline": [
    "1969-12-09 - Birth of Bixente Jean Michel Lizarazu in France",
    "1989-08-xx - Start of 1989-1990 French First Division season",
    "1991-08-xx - Start of 1991-1992 French Second Division season",
    "1996-05-08 - 1995-1996 UEFA Cup Final",
    "1997-07-01 - Lizarazu joins Bayern Munich in the Bundesliga",
    "2001-05-23 - 2001 UEFA Champions League Final"
  ],
  "topic": "Club career progression and major achievements of Bixente Lizarazu across France and Germany, including French and German domestic leagues"
}

CRITICAL REQUIREMENTS:
- Include EVERY entity or event mentioned directly or indirectly in the text.
- Ensure all delimiters and quotes are correctly placed.
- Verify that your output is valid JSON.
- Use only the specified keys; do not add additional keys.
- Output a JSON object only (no variable assignments or equal signs).

Given any new text chunk, output exactly this JSON structure with all fields populated from the text.
 


This is the text you must parse and provide metadata for:
Document Title: "{doc_title}"
Document Content: {doc_content}
\end{lstlisting}

\section{Prompts for Agentic baseline}
\subsection{AMAZON}
\begin{lstlisting}[language=Prompt,caption={Neo4j prompt for schema-aware retrieval on Amazon graph}]
### Neo4j Graph Database (Write Cypher Queries Directly)
For ALL graph traversal, write Cypher queries directly. Do NOT use function calls for graph traversal.

All graph data is in Neo4j. You must:
1) Understand the question: which entities and relationships are involved?
2) Plan the traversal: which node labels, relationships, and constraints are needed?
3) Write Cypher: output ONE Cypher query that returns the answer candidates.

Neo4j graph schema:
(Product)-[:HAS_REVIEW]->(Review)           Products have customer reviews
(Product)-[:ALSO_BUY]-(Product)             Co-purchase recommendations
(Product)-[:ALSO_VIEW]-(Product)            Co-view recommendations

Product properties (selected):
- p.nodeId (int, unique)
- p.asin (str, unique)              Amazon Standard Identification Number
- p.title (str)
- p.brand (str)
- p.price (str)                     String format: "$11.80"
- p.color (str/list)
- p.globalCategory (str)            High-level category
- p.category (list[str])            Specific categories
- p.feature (list[str])             Product features/specifications
- p.description (list[str])         Product descriptions
- p.details (str)                   JSON-encoded specifications
- p.rank (str)                      Sales rank

Review properties (selected):
- r.nodeId (int, unique)
- r.asin (str)                      Product ASIN
- r.reviewerID (str)
- r.overall (float)                 Star rating (1.0 to 5.0)
- r.reviewText (str)
- r.summary (str)                   Review title
- r.verified (bool)                 Verified purchase
- r.reviewTime (str)                Review date
- r.vote (str)                      Helpful vote count
- r.style (str)                     Product variant info (JSON)

Cypher examples:

Example 1: Products by brand in a category
MATCH (p:Product)
WHERE toLower(p.brand) CONTAINS 'nike'
RETURN p.nodeId

Example 2: Products with specific features
MATCH (p:Product)
WHERE toLower(p.title) CONTAINS 'backpack'
  AND (toLower(p.feature) CONTAINS 'waterproof' 
       OR toLower(p.description) CONTAINS 'waterproof')
RETURN p.nodeId

Example 3: Products with specific rated reviews, sorted by average rating
MATCH (p:Product)-[:HAS_REVIEW]->(r:Review)
WHERE toLower(p.title) CONTAINS 'baseball cap'
  AND r.overall >= 4.0
RETURN p.nodeId, avg(r.overall) AS avgRating, count(r) AS reviewCount
ORDER BY avgRating DESC, reviewCount DESC

Example 4: Products under a price threshold
MATCH (p:Product)
WHERE toLower(p.title) CONTAINS 'knife'
  AND toFloat(substring(p.price, 1)) < 20.0
RETURN p.nodeId

Example 5: Recommendations based on co-purchase
MATCH (bought:Product)-[:ALSO_BUY]-(p:Product)
WHERE bought.asin = '0000032042'
  AND toLower(p.title) CONTAINS 'accessories'
RETURN DISTINCT p.nodeId
\end{lstlisting}

\subsection{MAG}
\begin{lstlisting}[language=Prompt,caption={Neo4j prompt for schema-aware retrieval}]
### Neo4j Graph Database (Write Cypher Queries Directly)
For ALL graph traversal, write Cypher queries directly. Do NOT use function calls for graph traversal.

All graph data is in Neo4j. You must:
1) Understand the question: which entities and relationships are involved?
2) Plan the traversal: which node labels, relationships, and constraints are needed?
3) Write Cypher: output ONE Cypher query that returns the answer candidates.

Neo4j graph schema:
(Author)-[:AUTHORED]->(Paper)                 Authors write papers
(Paper)-[:CITES]->(Paper)                     Citation network
(Paper)-[:HAS_FIELD]->(Field)                 Fields of study
(Author)-[:AFFILIATED_WITH]->(Institution)    Author affiliations

Paper properties (selected):
- p.paperId (int, unique)
- p.title (str)
- p.abstract (str)
- p.year (int)
- p.date (str)  Use p.date for exact date constraints (e.g., '2016-02-11')
- p.journalDisplayName (str)
- p.docType (str)
- p.paperCitationCount (int)  Use for "most cited" constraints

Author properties (selected):
- a.authorId (int, unique)
- a.name (str)
- a.displayName (str)

Field properties (selected):
- f.fieldId (int, unique)
- f.name (str)

Institution properties (selected):
- i.institutionId (int, unique)
- i.name (str)

Cypher examples:

Example 1: Papers by an author in a given year
MATCH (:Author {authorId: 324400})-[:AUTHORED]->(p:Paper)
WHERE p.year = 2017
RETURN p.paperId

Example 2: Papers by exact date (use p.date, not only p.year)
MATCH (p:Paper)
WHERE p.date = '2016-02-11'
RETURN p.paperId

Example 3: Most cited paper matching a textual constraint
MATCH (p:Paper)
WHERE toLower(p.title) CONTAINS toLower('graph retrieval')
   OR toLower(p.abstract) CONTAINS toLower('graph retrieval')
RETURN p.paperId, p.paperCitationCount
ORDER BY p.paperCitationCount DESC
LIMIT 1
\end{lstlisting}
\subsection{PRIME}
\begin{lstlisting}[language=Prompt,caption={Neo4j prompt for schema-aware retrieval on PRIME graph}]
### Neo4j Graph Database (Write Cypher Queries Directly)
For ALL graph traversal, write Cypher queries directly. Do NOT use function calls for graph traversal.

All graph data is in Neo4j. You must:
1) Understand the question: which entities and relationships are involved?
2) Plan the traversal: which node labels, relationships, and constraints are needed?
3) Write Cypher: output ONE Cypher query that returns the answer candidates.

Neo4j graph schema:
(Gene)-[:INTERACTS_WITH_PROTEIN]-(Gene)              Protein-protein interactions
(Drug)-[:TARGETS]->(Gene)                            Drug targets gene/protein
(Drug)-[:INDICATED_FOR]->(Disease)                   Drug treats disease
(Drug)-[:CONTRAINDICATED_FOR]->(Disease)             Drug should NOT be used
(Drug)-[:SYNERGISTIC_WITH]-(Drug)                    Drug synergy
(Drug)-[:HAS_SIDE_EFFECT]->(Phenotype)               Drug side effects
(Gene)-[:ASSOCIATED_WITH]-(Disease)                  Gene-disease associations
(Gene)-[:EXPRESSED_IN]->(Anatomy)                    Gene expression in tissue
(Gene)-[:INTERACTS_WITH]->(Pathway)                  Gene in pathway
(Gene)-[:INTERACTS_WITH]->(MolecularFunction)        Gene function
(Gene)-[:INTERACTS_WITH]->(BiologicalProcess)        Gene in process
(Disease)-[:PHENOTYPE_PRESENT]->(Phenotype)          Disease symptoms
(Disease)-[:LINKED_TO]->(Exposure)                   Disease environmental links

Node properties (selected):
Gene:
- g.nodeId (int, unique)
- g.name (str)
- g.sourceId (str)
- g.detailsJson (str)               Contains: summary, aliases, location

Disease:
- d.nodeId (int, unique)
- d.name (str)
- d.sourceId (str)
- d.detailsJson (str)               Contains: definition, symptoms

Drug:
- drug.nodeId (int, unique)
- drug.name (str)
- drug.sourceId (str)
- drug.detailsJson (str)            Contains: mechanism, indication

Phenotype:
- p.nodeId (int, unique)
- p.name (str)

Pathway:
- pw.nodeId (int, unique)
- pw.name (str)
- pw.stId (str)                     Reactome stable ID
- pw.detailsJson (str)

Anatomy:
- a.nodeId (int, unique)
- a.name (str)

MolecularFunction, BiologicalProcess, CellularComponent:
- nodeId, name (standard properties)

Cypher examples:

Example 1: Disease with multiple phenotypes
MATCH (d:Disease)-[:PHENOTYPE_PRESENT]->(p:Phenotype)
WHERE p.name IN ['pharyngitis', 'chemosis']
RETURN DISTINCT d.nodeId, d.name

Example 2: Drug for a disease
MATCH (drug:Drug)-[:INDICATED_FOR]->(d:Disease)
WHERE d.name = 'sclerosing cholangitis'
RETURN drug.nodeId, drug.name

Example 3: Gene with protein-protein interactions
MATCH (g:Gene)-[:INTERACTS_WITH_PROTEIN]-(g2:Gene)
WHERE g2.name IN ['hbq1', 'sirt5']
RETURN DISTINCT g.nodeId, g.name

Example 4: Gene in pathway and biological process
MATCH (g:Gene)-[:INTERACTS_WITH]->(bp:BiologicalProcess)
MATCH (g)-[:INTERACTS_WITH]->(pw:Pathway)
WHERE bp.name = 'cellular response to manganese ion'
  AND toLower(pw.name) CONTAINS 'atp'
RETURN g.nodeId, g.name

Example 5: Drug with target and contraindication
MATCH (drug:Drug)-[:TARGETS]->(g:Gene)
MATCH (drug)-[:CONTRAINDICATED_FOR]->(d:Disease)
WHERE g.name = 'ccr5'
  AND d.name = 'gout'
RETURN drug.nodeId, drug.name
\end{lstlisting}

\end{document}